\documentclass[useAMS,usenatbib,times]{mn2e}

\usepackage[]{graphicx}
\usepackage{graphicx}
\usepackage{cite}
\usepackage{calc}
\usepackage[fleqn]{amsmath}
\usepackage{natbib}
\usepackage{float}
\usepackage{amssymb}

\title[The properties of energetically unbound stars in stellar clusters]{The properties of energetically unbound stars in stellar clusters}
\author[I. Claydon, M. Gieles and A. Zocchi]{Ian Claydon\thanks{E-mail:
i.claydon@surrey.ac.uk (IC); m.gieles@surrey.ac.uk (MG); a.zocchi@surrey.ac.uk (AZ)}, Mark
Gieles and Alice Zocchi\\
Department of Physics, University of Surrey, Guildford, GU2 7XH, UK.\\}
\begin{document}

\date{}

\pagerange{\pageref{firstpage}--\pageref{lastpage}} \pubyear{2016}

\maketitle

\label{firstpage}

\begin{abstract}
Several Milky Way star clusters show a roughly flat velocity dispersion profile at large radii, which is not expected from models with a tidal cut-off energy. Possible explanations for this excess velocity include: the effects of a dark matter halo, modified gravity theories and energetically unbound stars inside of clusters. These stars are known as potential escapers (PEs) and can exist indefinitely within clusters which are on circular orbits. Through a series of $N$-body simulations of star cluster systems, where we vary the galactic potential, orbital eccentricity and stellar mass function, we investigate the properties of the PEs and their effects on the kinematics. We derive a prediction for the scaling of the velocity dispersion at the Jacobi surface due to PEs, as a function of cluster mass, angular velocity of the cluster orbit, and slope of the mass profile of the host galaxy. We see a tentative signal of the mass and orbital velocity dependence in kinematic data of globular clusters from literature. We also find that the fraction of PEs depends sensitively on the galactic mass profile, reaching as high as 40\% in the cusp of a Navarro-Frenk-White profile and as the velocity anisotropy also depends on the slope of the galactic mass profile, we conclude that PEs provide an independent way of inferring the properties of the dark matter mass profile at the galactic radius of (globular) clusters in the \textit{Gaia} era.
\end{abstract}

\begin{keywords}
galaxies: kinematics and dynamics -– galaxies: star clusters –- methods:
analytical –- methods: $N$-body simulations.
\end{keywords}

\section{Introduction}
Investigations into the behaviour of stars in globular clusters (GCs) have unearthed peculiarities that are not consistent with the expected behaviour of bound stars. These include extended structure surrounding clusters (\citealt{Grillmair1995}; \citealt{Kuzma2016}) and unusual surface density profiles (\citealt{Cote2002}, \citealt{Carraro2009}; \citealt*{Kupper2011}), individual stars with velocities near or above the escape velocity (\citealt*{Meylan1991}; \citealt{Lutzgendorf2012}), and a flattening of the velocity dispersion profile at large radii.

This flattening has been observed in an increasing number of clusters (\citealt{Drukier1998}; \citealt{Scarpa2007}; \citealt{Lane2010}), although there are many cases where self-consistent models (\citealt{King1966}; \citealt{Wilson1975}) have accurately fit the observed velocity dispersion profile of Milky Way and local group clusters (\citealt{McLaughlin2005}; \citealt{Barmby2009}). It is not understood why some clusters show this feature and others do not, or how many clusters would be expected to display it.

Attempts to explain the flattening of the dispersion have ranged from the effects of extra-tidal stars to deviations from Newtonian gravity \citep{Scarpa2007}. In modified Newtonian dynamics (MOND, \citealt{Milgrom1983}) there is a transition into this regime from Newtonian dynamics if both the acceleration of the GC around the galaxy and the internal acceleration of stars fall below a threshold acceleration, which can correspond to the radial position where the velocity dispersion profiles begin to flatten \citep*{Hernandez2013}. Alternatively the $\Lambda$ cold dark matter model, $\Lambda$CDM, and the hierarchical merger scenario for galaxy formation suggest that GCs formed in dark matter halos (\citealt{Peebles1984}; \citealt{Kravtsov2005}). Although internal effects expel DM from inside of clusters \citep{Baumgardt2008} and tidal interactions would possibly strip the DM halo (\citealt{Moore1996}; \citealt{Mashchenko2005}), GCs on large galactocentric orbits could still contain this DM component, which would then interact gravitationally with stars in the cluster and increase their velocity dispersion (\citealt{Ibata2013}).

However \citet{Kupper2010}, hereafter K10, showed that a flattening of the velocity dispersion profile occurs in simulations using purely Newtonian dynamics. This is due to the effect of potential escapers (PEs), which are stars that orbit inside of GCs but with an energy above the critical energy required for escape (\citealt{Fukushige&Heggie2000}; from now on FH00). If in models of cluster evolution the tidal truncation is approximated as an energy truncation at the critical energy, then the lifetimes are proportional to the half-mass relaxation time $t_{\rm rh}$, because stars gain energy on a relaxation time. FH00 noted that if a tidal field is included the lifetimes show a weaker dependence on $t_{\rm rh}$. They found the cause to be a population of PEs which increases the dissolution time, $t_{\rm diss}$; this effect is more important for simulations with a lower number of stars, $N$.

\citet{Baumgardt2001}, hereafter B01, showed with a model of the PEs energy distribution that this delayed escape leads to a scaling of the lifetime of a cluster with $t_{\rm rh}^{3/4}$. A constant fraction of stars are scattered above the critical energy each $t_{\rm rh}$ \citep{Ambartsumian1985}, but they do not escape instantaneously, and it is possible that some can be on stable orbits if the cluster is on a circular orbit \citep{Henon1969}. Stars that gain a large energy kick from a single interaction can escape isotropically, however the majority of stars gain energy gradually via many encounters causing them to drift into the PE regime. These PEs can then only escape via narrow apertures around the Lagrangian points (FH00). For circular orbits, the Lagrangian points L1 and L2 are along a line defined by connecting the centre of the cluster to that of the galaxy, where the radial derivative of the total potential (the sum of the cluster potential, the tidal potential and centrifugal potential) is zero. It is also the furthest distance from the cluster centre of the last closed equipotential surface, or Jacobi surface (see e.g. Section 3.3 of \citealt{Binney1987}).

B01 found the scaling of the lifetimes with $t_{\rm rh}^{3/4}$ to be consistent with direct $N$-body models of star clusters orbiting in a point mass galactic potential. \citet{Tanikawa2010}, hereafter TF10, then studied the dynamical evolution of clusters in galaxies with different (power-law) density profiles with direct $N$-body simulations and confirmed the $t_{\rm rh}^{3/4}$ scaling of the lifetimes for clusters that are initially Roche-filling. They also showed that for clusters with the same $N$ and tidal radius, orbiting in different galactic potentials, those with the highest angular frequency (i.e. moving in flatter density profiles) live longest. For clusters orbiting in flatter galactic density profiles, the Jacobi surface is compressed (for the same Jacobi radius, $r_{\rm J}$), resulting in smaller escape annuli and therefore a larger $t_{\rm diss}$ (\citealt*{Renaud2011}; hereafter R11). This is contrary to  what is found for clusters on different orbits in a given potential, because in that case $t_{\rm diss} \propto 1/\Omega$, where $\Omega$ is the angular velocity of the cluster orbit about the galaxy centre.

Measurements of the kinematics of stars within globular clusters are mostly based on line-of-sight velocities. However, to properly characterise the velocity dispersion anisotropy and rotation of these systems, proper motion data are required. Various proper motion measurements have recently become available including observations using the Hubble Space Telescope (HST) (\citealt{Bellini2014}; \citealt{Watkins2015}), and the first data release (DR1) of the ESA \textit{Gaia} mission \citep{Gaia1}. DR1 provided proper motions of many field stars in the Milky Way and also included open cluster stars, and future releases will provide proper motions of stars in the outer regions of GCs. Therefore, understanding the effects of PEs on the kinematics is paramount to correctly interpreting the new data, as current models have been shown to still have large biases when comparing to projected data from simulations (\citealt{Shanahan2015}; \citealt{Sollima2015}), and will also help to develop a prescription for including their effects in a self-consistent model.

The focus of this study is to use a series of simulations to investigate the properties of PEs, including their spatial and energy distribution, their kinematics and their effect on the kinematics of the cluster as a whole. We do this to determine if there are any aspects of the PEs which could be used to observationally constrain the external Galactic potential, or if there are observable features of PEs which can be used to discriminate between alternative predictions proposed by MOND and DM theories.

The paper is organized as follows. In Section 2 we describe how the simulations were set up and what initial conditions were chosen. Section 3 investigates the amount of PEs that exist in the simulations, and their distribution and dynamics. In Section 4 we derive a prediction for the velocity dispersion at the Jacobi surface and compare this to simulations and observational data. Finally we present our conclusions in Section 5.

\section{Description of the Simulations}
 All simulations were run using \texttt{\small{NBODY6TT}} \citep{Renaud2015}, a modified version of the direct $N$-body integrator \texttt{\small{NBODY6}} \citep{Aarseth2003} optimised for use with GPUs \citep{Nitadori2012}. \texttt{\small{NBODY6TT}} (mode B) allows any functional input for the galactic potential and avoids a linearised approximation of the tidal forces. We considered power-law mass profiles for our galactic potential, using the notation from \citet*{Innanen1983} and their equation A2 for the mass enclosed within a distance from the centre of the galaxy $R_{\rm{g}}$,
\begin{equation}
M(<R_{\rm{g}}) = M_{\rm 0}\left(\frac{R_{\rm{g}}}{R_{\rm 0}}\right)^{\lambda},
\label{eq:Gmass}
\end{equation}
where $M_{\rm 0}$ and $R_{\rm 0}$ are scale-factors. From this they obtain the potential in their equations A11 and A12
\begin{equation}
\displaystyle
  \phi_{\rm g}(R_{\rm{g}})=\begin{cases}
    \displaystyle\frac{GM_{\rm 0}}{(\lambda-1)R_{\rm 0}}\left[\left(\frac{R_{\rm{g}}}{R_{\rm 0}}\right)^{\lambda-1}-1\right], & \text{if $\lambda > 0$, $\lambda \neq 1$ }.\\
    \displaystyle\frac{GM_{\rm{0}}}{R_{\rm 0}}\ln{\frac{R_{\rm{g}}}{R_{\rm 0}}}, & \text{if $\lambda = 1$}.
   \label{eqn:phi}
  \end{cases}
\end{equation}

We consider 3 specific cases, using $\lambda$ =0, 1 and 2 which correspond to a point mass, singular isothermal sphere, and a $1/R_{\rm g}$ density profile \citep*[i.e. the density profile within the scale radius of a Navarro-Frenk-White profile;][]{Navarro1996}, respectively. In each potential we simulate clusters with an initial number of stars $N_0$=16384 of the same mass, or with masses distributed according to the \citet{Kroupa2001} mass function between $0.1 M_{\odot}$ and  $1\, M_{\rm\odot}$. We also vary the eccentricities of the orbit, using $\epsilon$= 0, 0.25, 0.5 and 0.75. The equations of motion are solved in a non-rotating reference frame that orbits the galactic centre with the centre of mass of the cluster initially in the origin. For the analysis of the circular orbits we move the data to a corotating reference frame, where the $x$-axis joins the centre of the cluster and the galaxy, which is always located at $-R_{\rm{g}}$, and the $y$-axis is positive in the direction of the tangential component of the orbital velocity (FH00). This is required as it is only possible to explicitly identify PEs in the corotating frame using the Jacobi energy of the stars (see e.g. chapter 5 of \citealt{Spitzer1987}). For the eccentric orbits this is not possible and we therefore carry out our analysis in the non-rotating frame. In this paper we use $N$-body units \citep{Henon1971} where ${G}=1$, the initial total cluster mass $M_{\rm c}=1$ and initial total energy ${E_{\rm t}}=-1/4$.

\subsection{Input parameters}
\subsubsection{Circular orbits}
We set up the simulations such that the clusters on circular orbits in each potential have the same initial half-mass radius, ${r_{\rm hm}}$, and $r_{\rm{J}}$. The initial conditions correspond to a King model with $W_{0} = 5$ \citep{King1966}\footnote{We use \texttt{\small{LIMEPY}} (https://github.com/mgieles/limepy, \citealt{Gieles2015}) to generate the initial positions and velocities of the stars.}. However, as the King model describes spherical distribution of stars within the radius $r_{\rm t}$, using this in a tidal potential will introduce the presence of an initial population of PEs outside the Jacobi surface. This is because a Jacobi surface with $r_{\rm J}=r_{\rm t}$ is triaxial and flatter in the $y$ and $z$ axes than in the $x$-axis. Therefore we define our galactic potential such that $r_{\rm J}$ = 1.5$r_{\rm t}$: in this way the King model will sit within the Jacobi surface and have no initial PEs\footnote{There will still be a small $\lambda$-dependent population of PEs due to the $z$-axis of the Jacobi surface becoming increasingly flattened as $\lambda$ increases.}. The filling factor is then ${r_{\rm hm}}/{r_{\rm J}} \simeq 0.125$.  The Jacobi radius for circular orbits in a galaxy defined by equation~(\ref{eq:Gmass}) is
\begin{equation}
    r_{\rmn{J}} = \left[{G M_{\rmn{c}} \over (3-\lambda)\Omega^{2}}\right]^{1/3}
\label{eqn:tidalradius}
\end{equation}
from \citet{King1962}, where $\Omega$, for our galactic potential, is defined as
\begin{equation}
    \Omega^2 = \frac{GM(<R_{\rm{g}})}{R_{\rm{0}}^3} = 
\frac{GM_0}{R_0^3} \left(\frac{R_{\rm g}}{R_{\rm 0}}\right)^{\lambda}.
\end{equation}

\texttt{\small{NBODY6TT}} requires astrophysical units for the input values for the galactic potential and the orbit. We find values for $M_{0}$, $R_{0}$  and $R_{\rm g}$ in physical units that give us the desired $r_{\rm J}$. We keep $R_{\rm g}$ the same for the circular orbits in the different potentials and calculate the required circular velocity of the cluster as
\begin{equation}
    V_{\rm c}(R_{\rm g}) = \Omega R_{\rm g} = \sqrt{\frac{GM_0}{R_{\rm g}}}\left(\frac{R_{\rm g}}{R_0}\right)^{\lambda/2}.
    \label{eqn:Vc}
\end{equation}

\subsubsection{Reference frame}
To analyse the simulations in the corotating frame, the solid-body rotation of the cluster stars relative to the non-rotating frame needs to be removed. To find the velocity components in the corotating reference frame we use $\boldsymbol{v_{\rm cr}} = \boldsymbol{v_{\rm nr}} - \boldsymbol{v_{\rm sb}}$, where $\boldsymbol{v_{\rm cr}}$ and $\boldsymbol{v_{\rm nr}}$ are the velocity vectors in the corotating and nonrotating reference frames respectively, and $\boldsymbol{v_{\rm sb}}$ is the solid body rotation due to the choice of the frame, which corresponds to $(0,0,\Omega_{\varphi}\sqrt{x^2+y^2})$ in spherical coordinates, where $\varphi$ indicates the angle from the positive $x$-axis in the direction of the positive $y$-axis. The positions in the corotating frame are then found by rotating the Cartesian position vector in the nonrotating frame in the negative $\varphi$ direction across the angular offset between the two frames.

\subsubsection{Eccentric orbits}
\begin{table}

\begin{center}
    \caption{Input values for our series of simulations. Columns from left to right are: name of the simulation, orbital eccentricity $\epsilon$, apocentre radius $R_{\rm a}$, apocentre velocity $V_{\rm a}$ (both in $N$-body units), initial mass function IMF and slope of the enclosed mass of the galaxy, $\lambda$. All simulations have $N_0=16384$ particles.}
    \begin{tabular}{|l|l|l|l|l|l|l|}

    \hline
    Name & $\epsilon$ & $R_{\rm a}$ & $V_{\rm a}$ & IMF &  $\lambda$\\ \hline
    $\lambda$0$\epsilon$0 & 0.00 & 2494.8 & 85.10 & Delta &  0 \\
    $\lambda$0$\epsilon$0.25 & 0.25 & 3118.5 & 65.92 & Delta &  0 \\
    $\lambda$0$\epsilon$0.5 & 0.50 & 3742.2 & 49.13 & Delta &  0 \\
    $\lambda$0$\epsilon$0.75 & 0.75 & 4365.9 & 32.17 & Delta &  0 \\
    $\lambda$0$\epsilon$0K & 0.00 & 2494.8 & 85.10 & Kroupa &  0 \\
    $\lambda$0$\epsilon$0.25K & 0.25 & 3118.5 & 65.92 & Kroupa &  0 \\
    $\lambda$0$\epsilon$0.5K & 0.50 & 3742.2 & 49.13 & Kroupa &  0 \\
    $\lambda$0$\epsilon$0.75K & 0.75 & 4365.9 & 32.17 & Kroupa &  0 \\
    $\lambda$1$\epsilon$0 & 0.00 & 2494.8 & 104.23 & Delta &  1 \\
    $\lambda$1$\epsilon$0.25 & 0.25 & 3118.5 & 79.23 & Delta &  1 \\
    $\lambda$1$\epsilon$0.5 & 0.50 & 3742.2 & 54.63 & Delta &  1 \\
    $\lambda$1$\epsilon$0.75 & 0.75 & 4365.9 & 29.68 & Delta &  1 \\
    $\lambda$1$\epsilon$0K & 0.00 & 2494.8 & 104.23 & Kroupa &  1 \\
    $\lambda$1$\epsilon$0.25K & 0.25 & 3118.5 & 79.23 & Kroupa &  1 \\
    $\lambda$1$\epsilon$0.5K & 0.50 & 3742.2 & 54.63 & Kroupa &  1 \\
    $\lambda$1$\epsilon$0.75K & 0.75 & 4365.9 & 29.68 & Kroupa &  1 \\
    $\lambda$2$\epsilon$0 & 0.00 & 2494.8 & 147.40 & Delta &  2 \\
    $\lambda$2$\epsilon$0.25 & 0.25 & 3118.5 & 110.55 & Delta &  2 \\
    $\lambda$2$\epsilon$0.5 & 0.50 & 3742.2 & 73.70 & Delta &  2 \\
    $\lambda$2$\epsilon$0.75 & 0.75 & 4365.9 & 36.85 & Delta &  2 \\
    $\lambda$2$\epsilon$0K & 0.00 & 2494.8 & 147.40 & Kroupa &  2 \\
    $\lambda$2$\epsilon$0.25K & 0.25 & 3118.5 & 110.55 & Kroupa &  2 \\
    $\lambda$2$\epsilon$0.5K & 0.50 & 3742.2 & 73.70 & Kroupa &  2 \\
    $\lambda$2$\epsilon$0.75K & 0.75 & 4365.9 & 36.85 & Kroupa &  2 \\ \hline

    \end{tabular}
\end{center}
\end{table}

The kinematics and other properties such as the mass of the cluster vary over the course of an eccentric orbit. This is because $r_{\rm J}$ will expand and contract causing stars to effectively escape from the cluster and then be recaptured.

It can therefore be useful to approximate an eccentric orbit by a circular orbit that has the same dissolution time and mass evolution \citep{Cai2016}. This allows us to reduce these orbital variations by adopting an approximate $r_{\rm J}$, which we refer to as $r_{\rm J, circ}$, at any point in the eccentric orbit by using the angular velocity of a circular orbit with the same lifetime.

To achieve this, we set up our eccentric orbit simulations with the same semi-major axis of the orbit, $a$, because then the lifetime is (to first order) independent  of eccentricity (\citealt{Cai2016}; Bar-or et al. in prep). The semi-major axes of the eccentric and circular orbits are
\begin{equation}
   a =\begin{cases}
   (R_{\rm a} + R_{\rm p})/2, & \text{if $\epsilon>0$}\\
   R_{\rm g}, & \text{if $\epsilon=0$},
   \end{cases}
\end{equation}
where $R_{\rm a}$ and $R_{\rm p}$ are the apocentre and pericentre distances respectively, and $\epsilon$ is the eccentricity of the orbit. By using the relation $R_{\rm p} = R_{\rm a}(1-\epsilon)/(1+\epsilon)$, we find
\begin{equation}
    R_{\rm a} = (1+\epsilon)a.
    \label{eqn:Ra}
\end{equation}
This gives a simple relation for the apocentre value depending only on the eccentricity and is independent of the potential.

To calculate the required initial apocentre velocity for the eccentric orbits, we use conservation of $E_{\rm t}$, and angular momentum of the orbit, $J$:
\begin{equation}
	\begin{split}
		E_{\rm t} &= E_{\rm a} = E_{\rm p}\\
		&= 0.5V_{\rm a}^2 + \phi_{\rm g}(R_{\rm a}) = 0.5V_{\rm p}^2 + \phi_{\rm g}(R_{\rm p}),
	\end{split}
\label{eqn:consE}
\end{equation}
and
\begin{equation}
	\begin{split}
		J &= J_{\rm a} = J_{\rm p} \\
		&= R_{\rm a}V_{\rm a} = R_{\rm p}V_{\rm p},
	\end{split}
\label{eqn:consJ}
\end{equation}
where the subscripts again refer to apocentre and pericentre. By substituting $V_{\rm p}$ in equation~(\ref{eqn:consE}) by using equation~(\ref{eqn:consJ}), we find
\begin{equation}
V_{\rm {a}}^2 = \frac{2\left[\phi(R_{\rm p}) - \phi(R_{\rm a})\right]}{1-\left({R_{\rm a}}/{R_{\rm p}}\right)^2}.
\label{eq:apoV}
\end{equation}
Then by using equations~(\ref{eqn:phi}),~(\ref{eqn:Vc}) and~(\ref{eqn:Ra}) we find the initial apocentre velocities for the eccentric orbits as
\begin{equation}
V_{\rm a}^2 =\begin{cases}
    \displaystyle \frac{2V_{\rm c}(R_{\rm a})^2}{(\lambda-1)} \left[\frac{\left(\frac{1-\epsilon}{1+\epsilon}\right)^{\lambda-1}-1}{1-\left(\frac{1+\epsilon}{1-\epsilon}\right)^2}\right], & \text{if $\lambda\neq1$}.\\
    \displaystyle V_{\rm c}(R_{\rm a})^2 \frac{2\ln{\left(\frac{1-\epsilon}{1+\epsilon}\right)}}{1-\left(\frac{1+\epsilon}{1-\epsilon}\right)^2}, & \text{if $\lambda=1$}.
    \end{cases}
\end{equation}

Table 1 shows the input values for all the simulations. The names of the simulations specify the values of $\lambda$, $\epsilon$ and the type of mass function.

\section{Properties of potential escapers}
\subsection{Definition and identification}
\begin{figure*}

    \centering
        \includegraphics[width=0.99\textwidth]{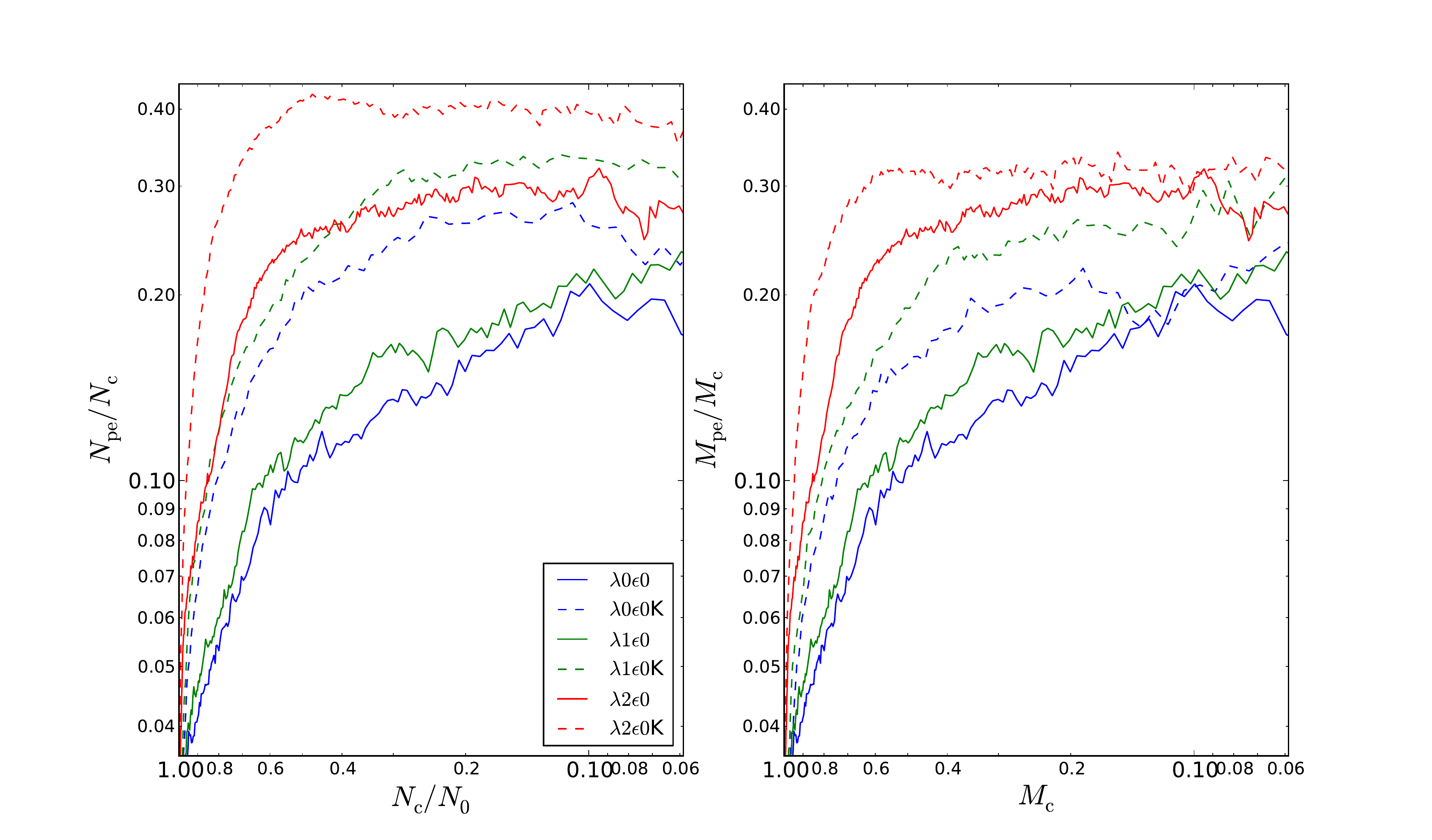}
    \caption{Ratio of the number of PEs to total number of stars remaining in the cluster (left) and the fraction of mass in PEs (right), for the circular orbits in each potential (each identified by a different colour, as indicated in the panel on the left plot) with equal-mass stars (solid lines) and Kroupa IMF (dashed lines).}
    
    \label{fig:Npe}

\end{figure*}
The Jacobi energy of a star is defined as (see e.g. page 2 of FH00)
\begin{equation}
E_{\rm J} = \frac{v^2}{2} + \phi_c + \frac{1}{2}\Omega^2\left[z^2 - (3-\lambda)x^2\right],
\label{eq:Ej}
\end{equation}
where we include the dependence on $\lambda$ (see the derivation in R11). The third term is a combination of the tidal and the centrifugal potentials when working in a corotating reference frame. 

We also define $\hat{E} = (E_{\rm J} - E_{\rm crit})/|E_{\rm crit}|$, where $E_{\rm crit} = -3GM_{\rm c}/2r_{\rm J}$ in a corotating reference frame. It is difficult to define exactly what constitutes a PE as there can be some stars with an energy above the critical energy on stable orbits inside the cluster \citep{Henon1969}, and others with apocentres outside of the Jacobi surface of the cluster. At any moment there will also be unbound stars that are in the process of isotropically escaping from the cluster but are still found within the Jacobi surface. To proceed we adopt the following working definition: \textit{PEs are stars inside a sphere of radius $r_{\rm J}$ that have $\hat{E} > 0$}. The maximum extent of the Jacobi surface on the $y$-axis is $(2/3)r_{\rm J}$ and along the $z$-axis the maximum point of the surface is $\lambda$-dependent, and is $\sim 0.638r_{\rm J},0.626r_{\rm J}$ and $0.596r_{\rm J}$ for $\lambda$=0, 1 and 2 respectively (see equation 14 of R11). This means that our definition of PEs using a sphere of radius $r_{\rm J}$ will include most of the stars which have the apocentre of their orbit outside of the Jacobi surface but will also include some unbound stars that have escaped from the Jacobi surface. A similar approximation is used when dealing with observational data, as a circular projected tidal surface is usually assumed.

\subsection{Properties and distribution}
\subsubsection{Fraction of PEs and mass distribution}

We begin our investigation by looking at the fraction of PEs relative to bound stars inside a sphere of radius $r_{\rm J}$. Fig.~\ref{fig:Npe} shows the ratio of the number of PEs to the total number of stars (left-hand panel) and the fraction of the total mass of stars in PEs (right-hand panel). Solid lines represent the simulations with equal-mass stars and dashed lines represent the simulations with a mass spectrum. The later stages of the $\lambda$0$\epsilon$0 simulation are consistent with the evolution found in B01, however there is a clear increase in the fraction when increasing $\lambda$. This increase is possibly due to the dependence of the escape time of individual stars, $t_{\rm e}$, on galactic potential: R11 and TF10 derived a $\lambda$ dependent $t_{\rm e}$ based on the flux of orbits out of the Lagrange points, finding  $t_{\rm e}(\lambda=2)/t_{\rm e}(\lambda=0)\sim$ 1.2 and 1.14 respectively\footnote{TF10 however found this ratio to be smaller than would be required from the differences in the dissolution times of their $N$-body simulations.}.

The number of PEs also increases when introducing a mass spectrum. The creation of PEs is due to many minor interactions with other stars and there is a constant amount of PEs created on the half-mass relaxation time-scale,
\begin{equation}
t_{\rm rh} \propto \frac{N_{\rm c}^{1/2}{r_{\rm hm}^{3/2}}}{\ln{\Lambda}{<m>^{1/2}\phi}}
\end{equation}
where $m$ is the mass of the individual stars, $<>$ indicates a mean, $\phi = < m^{5/2}>/<m>^{5/2}$, which equals 1 when the masses of the stars are equal \citep{Spitzer1971}, $\ln{\Lambda}$ is the Coulomb logarithm with $\Lambda=0.11N_{\rm c}$ \citep{Giersz1994} and $N_{\rm c}$ is the number of stars inside the cluster. Therefore, systems that have a spectrum of masses have a shorter $t_{\rm rh}$ resulting in a higher production rate of PEs compared to a system with equal-mass stars. Because the escape time is not dependent on the mass function, more PEs build up in the simulations of clusters with a spectrum of masses. This increasing fraction of PEs for higher $\lambda$ (i.e. galaxies with flatter density profiles) corroborates the 0.35 fraction found in \citet{Just2009}, where they used a Salpeter IMF \citep{Salpeter1995} and Miyamoto-Nagai disk for their galactic potential \citep{Miyamoto1975}. 

There is an initial phase of rapid PEs production where more PEs are produced than escape from the cluster. Although our initial value of $r_{\rm t}/r_{\rm J}$ avoided having any primordial PEs, there is a large amount of stars that are very close to the critical energy and therefore take less time to be scattered above it.  After this phase the gradient decreases, which is much more noticeable for the $\lambda$=2 and mass spectrum simulations, as the production and loss of PEs becomes closer to being balanced. In simulations with lower particle number (not shown in the figure), we found that by increasing the initial value of $r_{\rm t}/r_{\rm J}$ the same final fraction of PEs is reached, but there is a lower fraction relative to Fig.~\ref{fig:Npe} for much of the lifetime. There is also an $N$-dependence in the fraction of PEs (B01) which possibly reduces their effects in systems with larger particles, but our simulations are directly comparable to the size of open cluster-like systems.

\begin{figure}

    \centering
        \includegraphics[width=0.99\columnwidth]{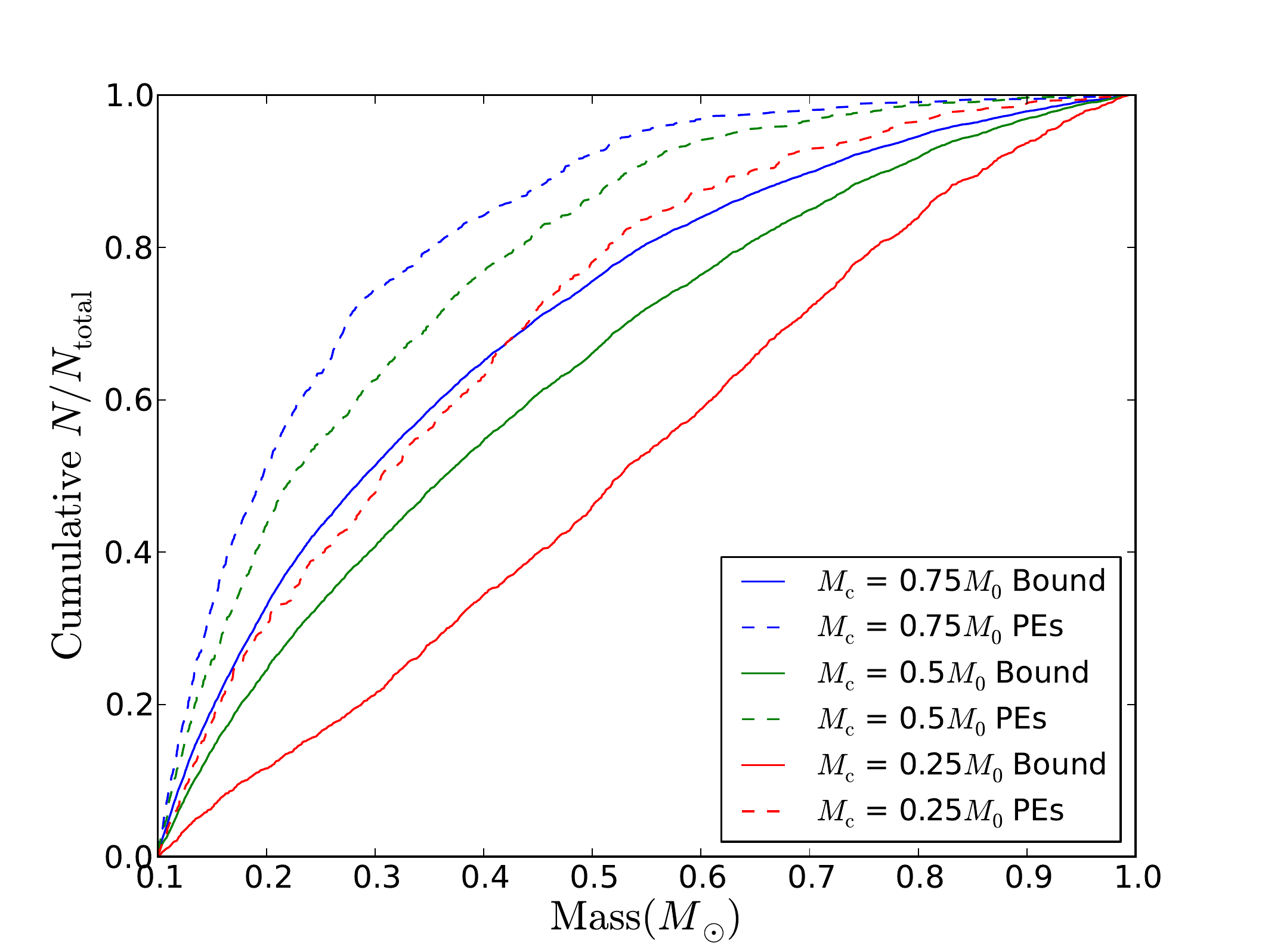}
    \caption{Cumulative mass functions for bound stars (solid lines) and PEs (dashed) at three different moments in the evolution of the $\lambda$1$\epsilon$0K simulation.}
    \label{fig:cummass}

\end{figure}

The right-hand panel of Fig.~\ref{fig:cummass} shows the fraction of mass in PEs which is lower than the number fraction for each of the models. This means that PEs are predominantly low mass, possibly as it is easier to scatter them above the critical energy. Fig.~\ref{fig:cummass} further investigates the mass of the PEs compared to the bound stars inside the cluster. We plot the cumulative fraction of stars as a function of mass of PEs (dashed) and bound stars (solid) at three snapshots when remaining mass is $0.75M_{0}$ (blue), $0.5M_{0}$ (green) and $0.25M_{0}$ (red), where $M_{0}$ is the initial mass. The PEs have a much higher fraction of stars with low mass, as expected. Even when the mass remaining is $0.25M_{0}$, over $40\%$ of the PEs are below $0.3M_{\odot}$, which means that a large amount of PEs may be below current observational limits. This could explain why the effects of PEs are ubiquitous in simulations yet the peculiarities in observations can vary.
\begin{figure}
    \centering
    \includegraphics[width=1\columnwidth]{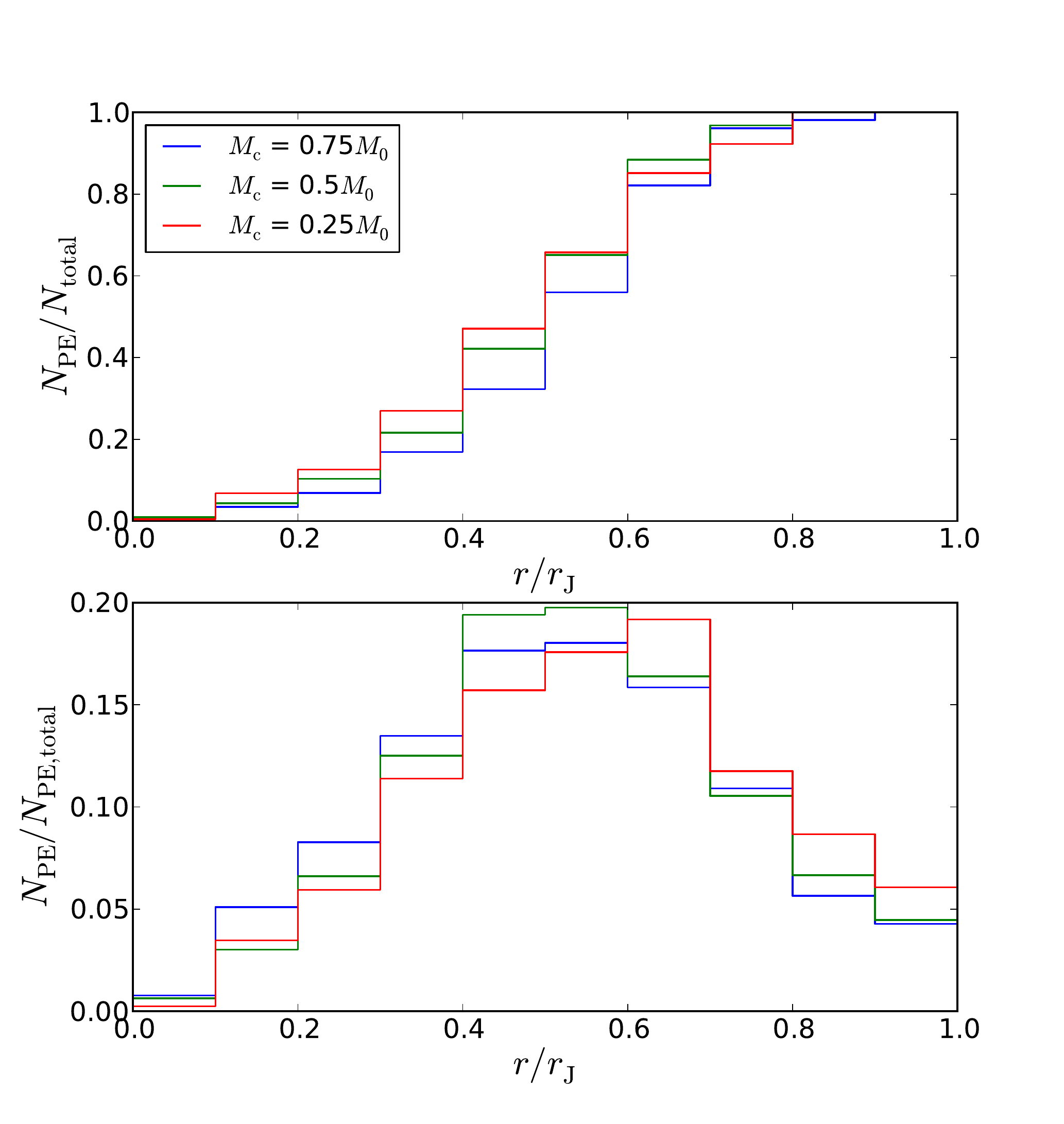}
    \caption{Top panel - Fraction of PEs to total number of stars in spherical bins. Bottom panel - Fractions of PEs in spherical bins to total number of PEs. Each panel shows three different moments through the lifetime of the $\lambda$1$\epsilon$0K simulation when mass of the cluster is 0.75, 0.5 and 0.25 of the initial cluster mass.}
    \label{fig:cum}
\end{figure}

\subsubsection{Spatial distributions}

The top panel of Fig.~\ref{fig:cum} represents the fraction of PEs to total stars in spherical bins of increasing radius, plotted at three points over the lifetime of the $\lambda$1$\epsilon$0K simulation. At all moments there is roughly an equal number of PEs and bound stars at $\sim$0.5$r_{\rm J}$ suggesting that the effect of PEs on the kinematics should be seen far into the cluster, as found in K10. Beyond this location the PEs dominate and beyond $\sim$0.8$r_{\rm J}$ approximately all stars are PEs, suggesting there are few bound stars that reach close to the Lagrange point, although there will be many PEs outside of the Jacobi surface in the outer spherical bins. The bottom panel shows the fraction of PEs to total number of PEs in spherical bins at that time. This quantity also peaks at around $\sim$0.5$r_{\rm J}$, and the location of this peak moves outwards slowly with time. The behaviour in the $\lambda$=1 simulation shown here is similar to the behaviour of the circular orbits in the other potentials. 

\begin{figure*}
    \centering
    \includegraphics[width=0.99\textwidth]{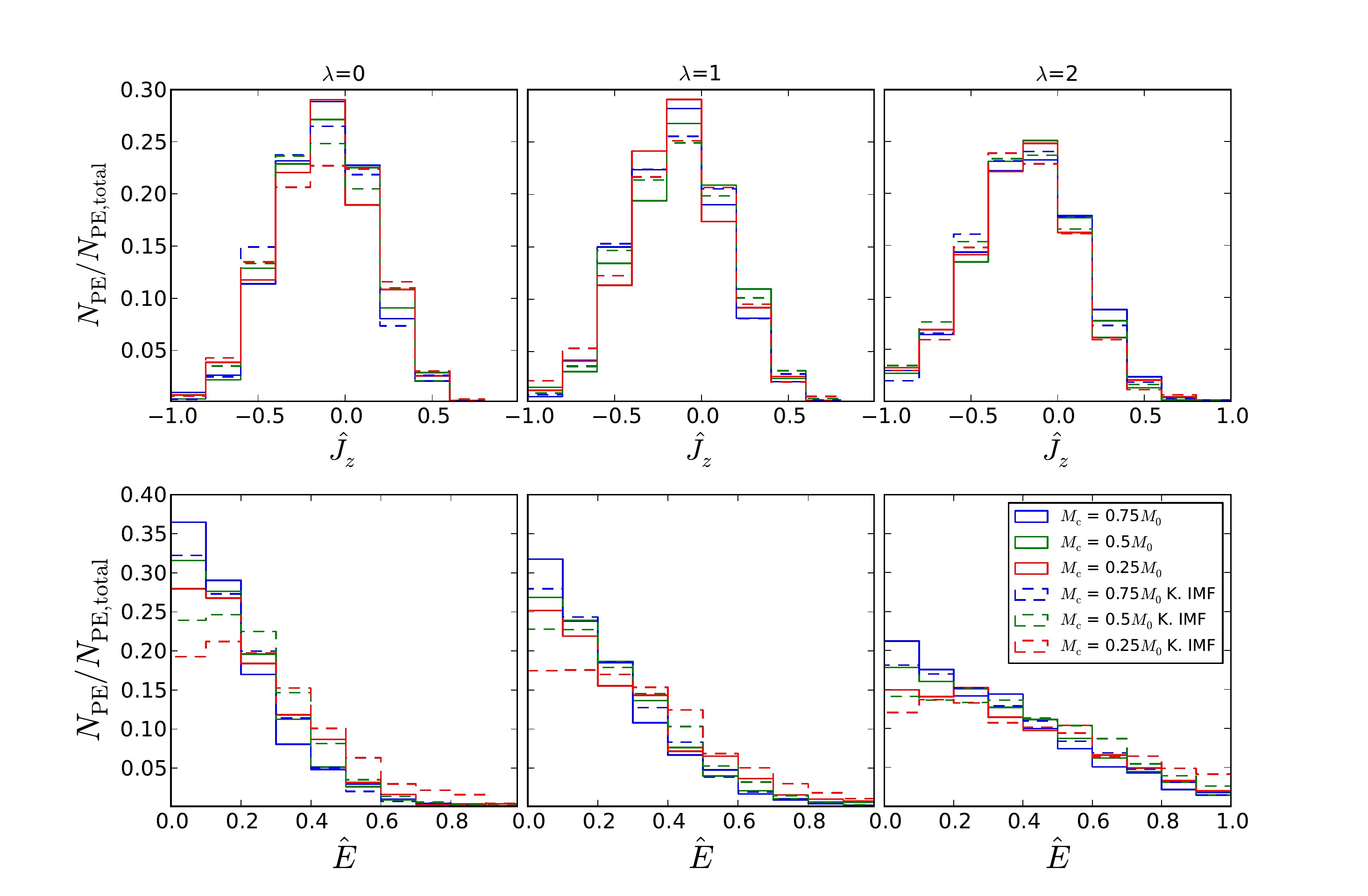}
    \caption{Histogram of $\hat{J_{z}}$ (top) and $\hat{E}$ (bottom) of PEs normalised to the total number of PEs. Solid lines are the equal-mass simulations, dashed lines are those with a Kroupa IMF. Both cases are shown at three different moments throughout the lifetime of the simulations at 0.75$M_0$, 0.5$M_0$ and 0.25$M_0$. Left-hand panels are $\lambda$=0, middle panels are $\lambda$=1 and right-hand panels are $\lambda$=2.}
    \label{fig:PE_ehat}
\end{figure*}


\subsubsection{Energy and angular momentum}

Fig.~\ref{fig:PE_ehat} shows dimensionless quantities of the energy and angular momentum of the PEs, scaled to properties of the cluster to determine if there is any variation in time of the PEs relative to the cluster. To do this we divide the $z$-component of the angular momentum by the angular momentum of a circular orbit at the Jacobi radius, $r_{\rm{J}}v_{\rm{c}}$, where $v_{\rm c}$ is the circular velocity of a fiducial star at the Jacobi radius, and call this quantity $\hat{J_{\rm{z}}}$ (top panels). For the energy we use $\hat{E}$ (bottom panels). Solid lines are the equal-mass clusters, dashed lines are the simulations using a Kroupa IMF. Both are displayed for three snapshots, when the mass remaining is 0.75$M_{\rm 0}$ (blue), 0.5$M_{\rm 0}$ (green) and 0.25$M_{\rm 0}$ (red). The panels from left to right represent the $\lambda$=0, 1 and 2 circular orbits respectively. There is minimal evolution in $\hat{J_{z}}$ for all simulations and there is little difference between the equal-mass and mass spectrum clusters. There is a negative bias which suggests a retrograde motion in the corotating reference frame.

The distribution in energy becomes wider with time (i.e. at lower $N$) for the clusters in each galactic potential, and this behaviour is more pronounced in the mass spectrum simulations. It is also evident that the distribution becomes wider with increasing $\lambda$, with a larger fraction of stars at higher energies.

By solving an equation similar to the Fokker-Planck equation, that considers the production, via diffusion, and escape of PEs, B01 introduced a model for the distribution $N(\hat{E})$ of PEs
\begin{equation}
N(\hat{E}) \propto \hat{E}^{1/2}K_{1/4}\left[\frac{1}{2}\left(\frac{t_{\rm rh}}{k_{1}t_{\rm esc}}\right)^{1/2}\hat{E}^2\right],
\label{eqn:Ne}
\end{equation}
where $K_{1/4}$ is a modified Bessel function, $t_{\rm esc}$ is the time for escape of a star with $\hat{E}=1$ and $k_{1}$ is a constant that corresponds to the fraction of mass scattered above $E_{\rm crit}$ over one $t_{\rm rh}$, the instantaneous half-mass relaxation time. 

\begin{figure}
    \centering
    \includegraphics[width=1\columnwidth]{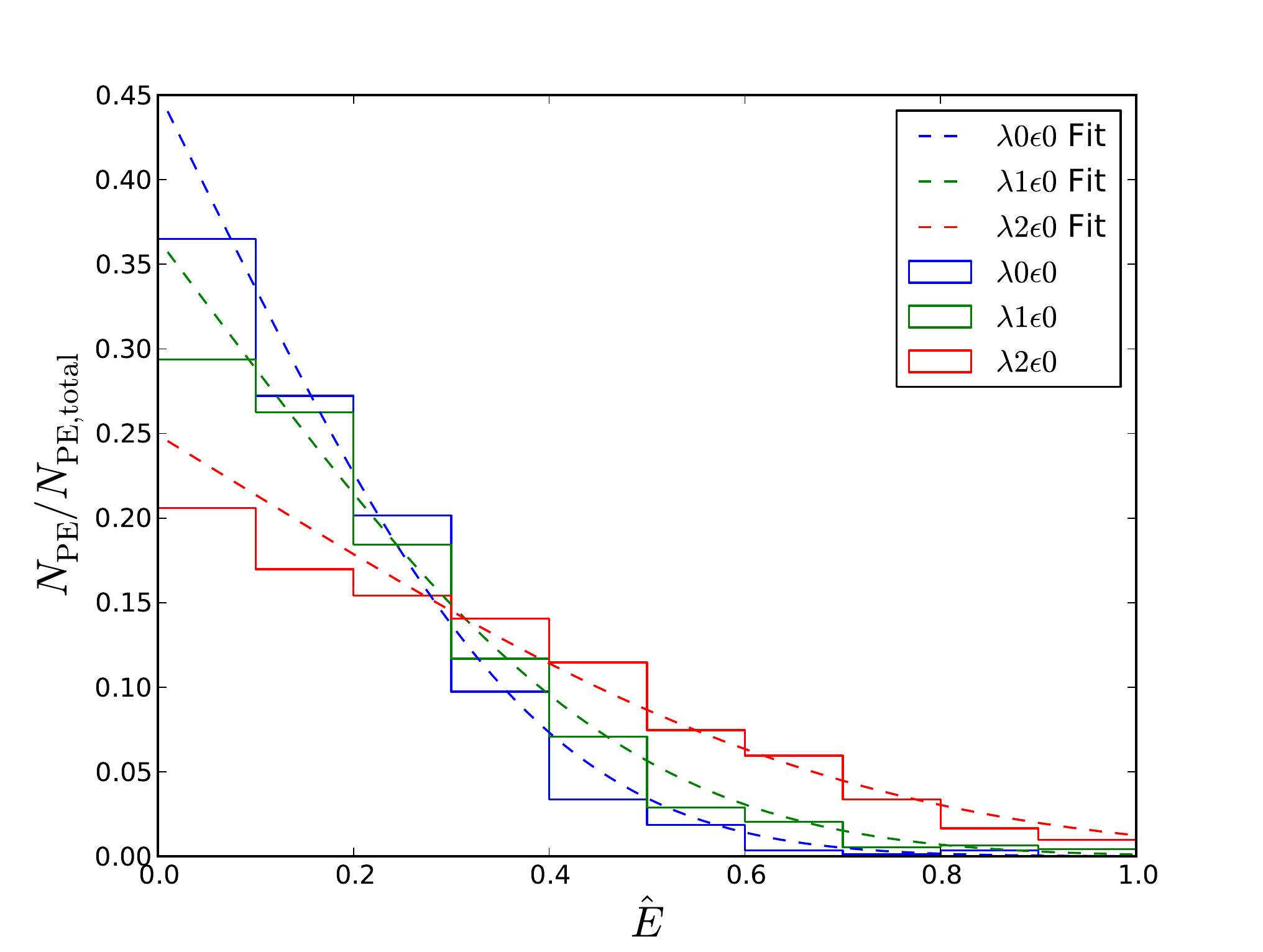}
    \caption{Fraction of PEs as a function of their energy at $M=0.5M_0$ in the $\lambda0\epsilon0$, $\lambda1\epsilon0$ and $\lambda2\epsilon0$ simulations. Blue, green and red lines are fit to each distribution.}
    \label{fig:PE_ehat_model}
\end{figure}

Fig.~\ref{fig:PE_ehat_model} shows the normalised $N(\hat{E})$ distribution for the $\lambda0\epsilon0$, $\lambda1\epsilon0$ and $\lambda2\epsilon0$ simulations (blue, green and red histograms respectively) when the clusters have a remaining mass of $0.5M_{\rm 0}$. 

We can express the half mass relaxation time as $t_{\rm rh} \propto (M_{\rm c}/<m>ln\Lambda)(r_{\rm hm}/r_{\rm J})^{3/2}(GM_{\rm c}/r_{\rm J}^3)^{1/2}$ and we consider the following expression (from FH00) for the escape time: $t_{\rm esc} \propto (GM_{\rm c}/r_{\rm J}^3)^{1/2} f(\lambda)$ where we have included a dependence on the galactic potential via $f(\lambda)$. We consider an empirical estimation of the function $f(\lambda) = [3/(3-\lambda)]^\alpha$ and by fitting on the distribution for each potential (blue, green and red lines in the figure), we find $\alpha \sim 1$.

This $\lambda$ dependence gives a variation in $t_{\rm esc}$ of $\sim$3 between $\lambda$=0 and $\lambda$=2, which is larger than the values found by R11 and TF10. However, our difference in dissolution times with $\lambda$ are consistent with the $N$-body simulations in TF10. It is important to note that $r_{\rm hm}/r_{\rm J}$ (and therefore $t_{\rm rh}$) also varies with $\lambda$: $r_{\rm hm}/r_{\rm J}$ will reduce to $\sim0.1$ at core collapse and then increase to 0.2 for $\lambda$=0 and 0.25 for $\lambda$=2.

This evolution of the energy and the variation with $\lambda$ can be used to derive an expression for the velocity dispersion at the Jacobi surface of a cluster, which we discuss further in Section 4.

\subsection{Dynamics of the potential escapers}
\subsubsection{Velocity dispersion}

We also explore the dynamics of the PEs and their effect on the kinematics of the cluster. The 1D velocity dispersion is calculated for each component of the velocity as
\begin{equation}
\sigma_{\rm 1D}^2 = <(v_{\rm 1D}^2 - <v_{\rm 1D}>^2)>.
\end{equation}
The 3D dispersion is then calculated for spherical coordinates, where $r$ is the radial component, $\theta$ is the angle from the positive $z$-axis and $\varphi$ is the angle measured from the $x$-axis in the $xy$ plane,
\begin{equation}
\sigma_{3D} = \sqrt{\sigma_{r}^2 + \sigma_{\theta}^2 + \sigma_{\varphi}^2}.
\end{equation}

\begin{figure}
    \centering
    \includegraphics[width=0.99\columnwidth]{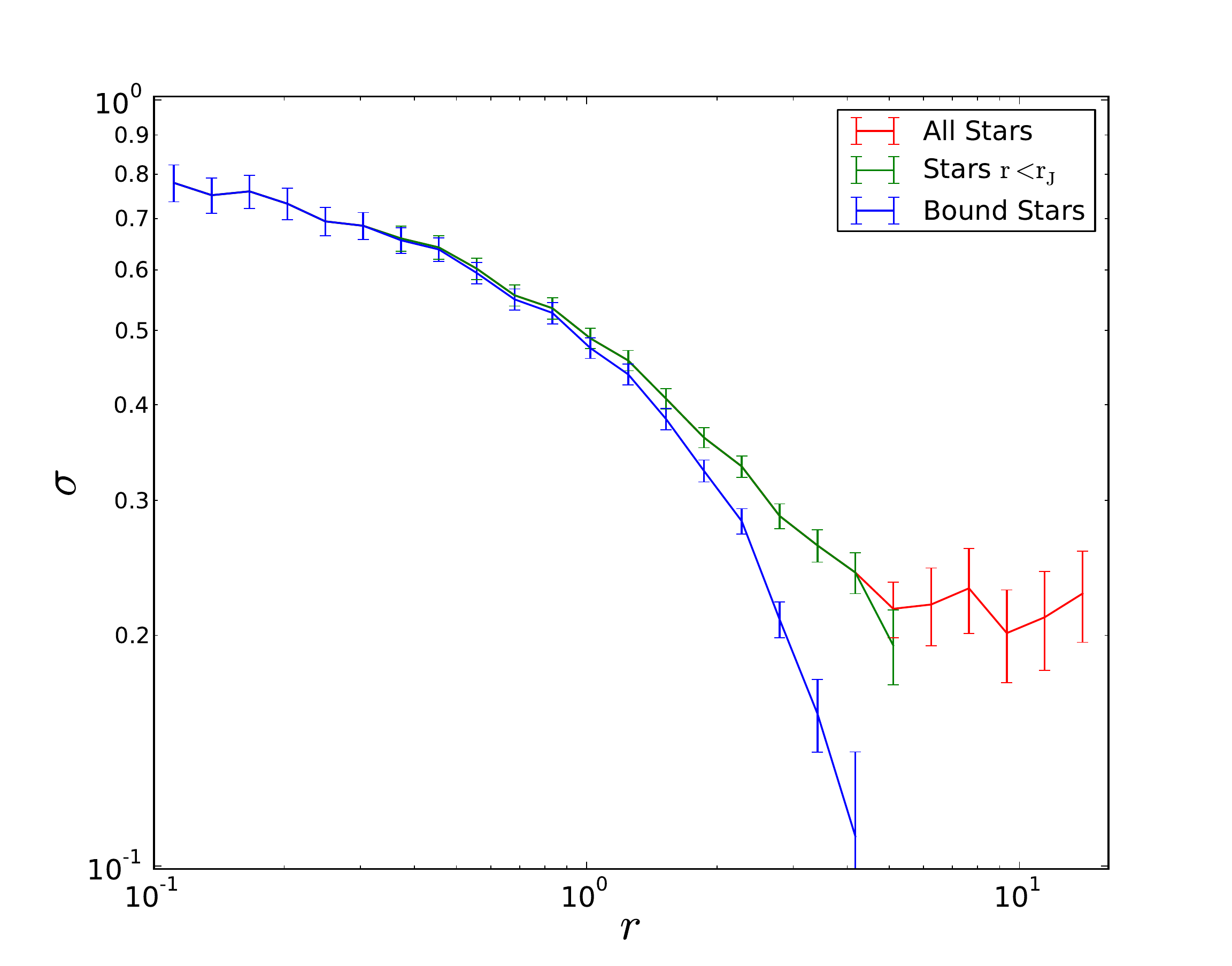}
    \caption{Radial profile of the dispersion of the $\lambda$1$\epsilon$0 simulation at 0.5 $M_0$. The blue lines represent stars within $r_{\rm J}$ with an energy below $E_{\rm crit}$, the green lines are all the stars within $r_{\rm J}$, and the red lines are all of the stars in the simulation.}
    \label{fig:dispprofile}
\end{figure}

Fig.~\ref{fig:dispprofile} shows the radial profiles of the $\sigma_{3D}$ in spherical bins for stars with $\hat{E} < 0$ (bound stars) in blue, all stars within $r_{\rm J}$ in green and all stars in red, for the $\lambda$1$\epsilon$0 simulation when the remaining mass is 0.5 $M_0$. The difference between the stars within $r_{\rm J}$ and the bound stars shows the effect of the PEs. The bound stars show a much sharper drop while the dispersion of all stars reduces less rapidly with distance from the centre. The difference between the PEs and the bound stars also increases with eccentricity of the orbit, as shown by K10 with numerical simulations. It can also be seen by taking projected quantities from simulations that the observational angle will also affect the velocity dispersion profile, as including stars belonging to the tidal tails will cause an increase in the dispersion. 

\begin{figure}
    \centering
    \includegraphics[width=0.99\columnwidth]{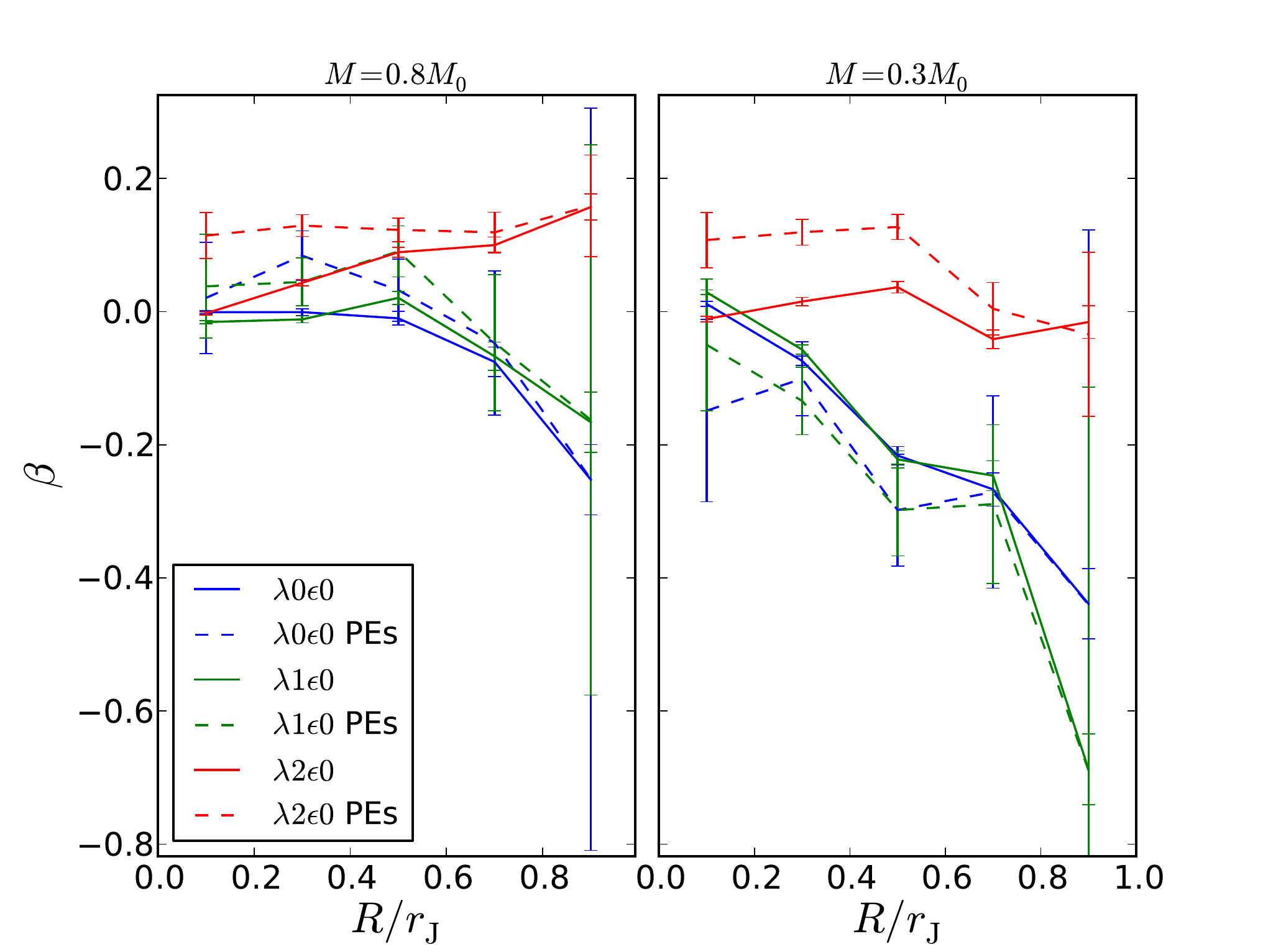}
    \caption{Radial profile for the anisotropy, $\beta$, in cylindrical bins of 0.2$r_{\rm J}$ width, for the $\lambda$0$\epsilon$0, $\lambda$1$\epsilon$0 and $\lambda$2$\epsilon$0 simulations. Left hand plot is the mean of two orbits around the time when the remaining mass is 0.8$M_0$, and the right plot is the same at a remaining mass of 0.3$M_0$.}
    \label{fig:Betapot}
\end{figure}

\begin{figure}
    \centering
    \includegraphics[width=0.99\columnwidth]{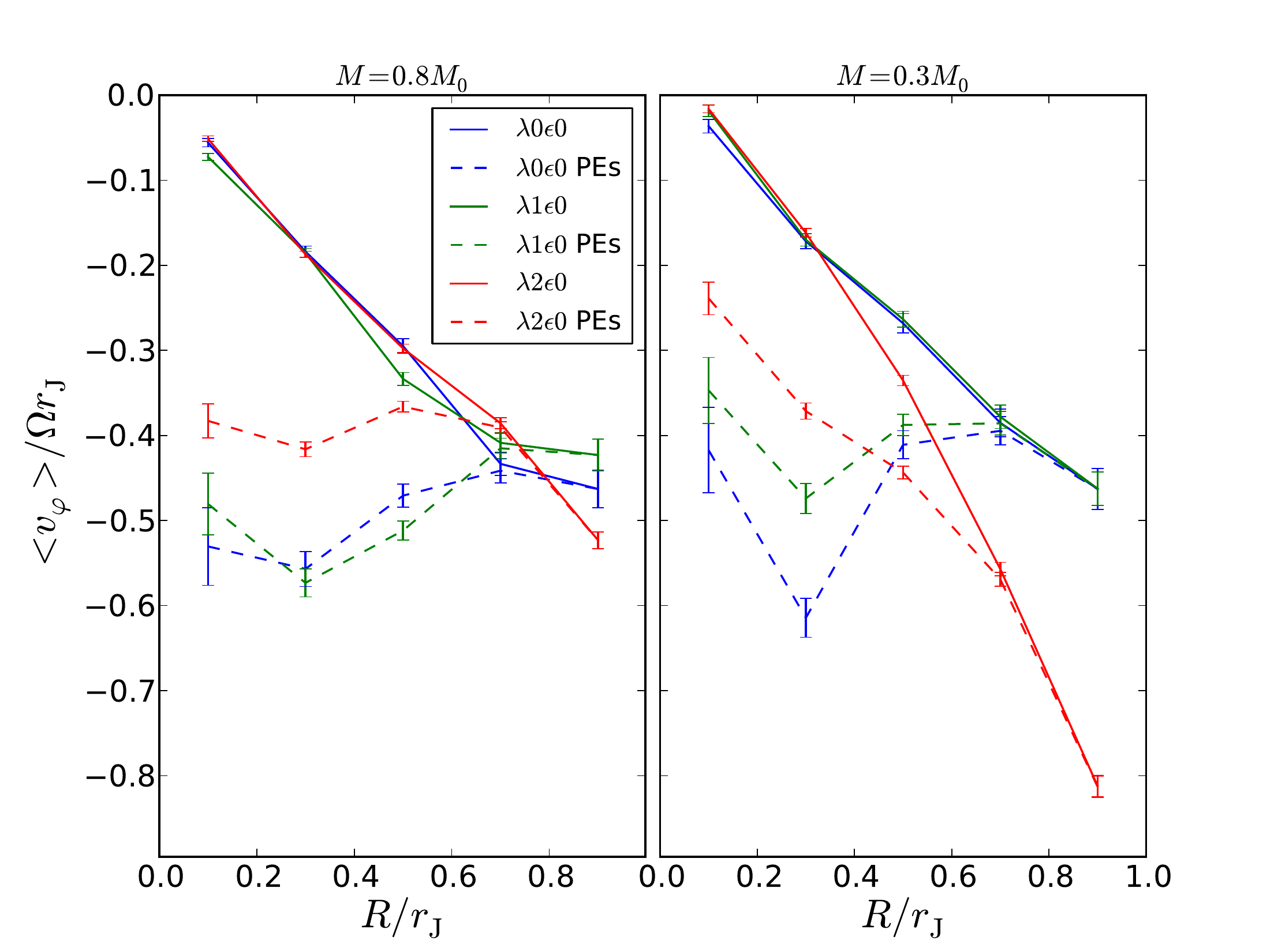}
    \caption{Radial $<v_{\varphi}>$ profile, normalised to $\Omega r_{\rm J}$, for the circular orbits in each potential using a corotating frame. Left hand plot is the mean of two orbits around the time when the remaining mass is 0.8$M_0$, and the right plot is the same at a remaining mass of 0.3$M_0$. Solid lines are all stars, dashed line are only the PEs.}
    \label{fig:vphi_norm}
\end{figure}

\subsubsection{Anisotropy of the dispersion and rotation}

To analyse the anisotropy of the dispersion in our simulations we use the $\beta$ parameter defined as
\begin{equation}
    \beta = 1 - \frac{\sigma_{\rm t}^2}{2\sigma_{\rm r}^2}
\end{equation}
where $\sigma_{\rm t}^2$=$\sigma_{\rm \theta}^2+\sigma_{\rm \varphi}^2$, $0<\beta\le1$ corresponds to radial anisotropy, $\beta<0$ to tangential anisotropy and $\beta=0$ to isotropy.

Fig.~\ref{fig:Betapot} shows the radial profile of $\beta$ for all the stars (solid lines) and only the PEs (dashed lines) in the $\lambda$0$\epsilon$0, $\lambda$1$\epsilon$0 and $\lambda$2$\epsilon$0 simulations at snapshots when the mass remaining is $0.8M_{\rm 0}$ (left-hand panel) and $0.3M_{\rm 0}$ (right-hand panel). We calculate $\beta$ in cylindrical bins, denoted by $R=\sqrt{x^2 + y^2}$, for each individual snapshot and take the mean for each bin over two orbits. We use cylindrical bins as the anisotropy for each bin is then the same in the corotating and nonrotating reference frames. The profiles in each potential are similar in the early snapshot (left-hand panel) where all are close to zero. The $\lambda$=0 and $\lambda$=1 simulations show some tangential anisotropy in the outer region, although the error bars are very large, whereas the $\lambda$=2 simulation appears to be isotropic, or slightly radially anisotropic. In the later snapshot a clearer difference between the potentials is visible with the $\lambda$=0 and $\lambda$=1 simulations developing tangential anisotropy, whereas the $\lambda$=2 simulation is isotropic. For all the potentials the bound stars are consistent with isotropy and the anisotropy that develops is contained mostly in the PEs. 

It has been shown that simulations of GCs with dense starting conditions develop radial anisotropy (\citealt{Sollima2015}; \citealt{Zocchi2016}) but those with larger initial $r_{\rm hm}/r_{\rm J}$, similar to our initial conditions, do not develop any radial anisotropy and instead show tangential anisotropy near the tidal radius \citep{Baumgardt2003}. This is thought to be due to the balance between the preferential production and preferential loss of radial orbits: two-body interactions predominantly scatter stars outwards on radial orbits, and these stars then escape more easily than those on other orbits (\citealt*{Takahashi97}, \citealt*{Tiongco2016}).

Therefore for dense initial conditions more stars are scattered outwards than can escape, which causes radial orbits to build up, but for extended clusters these radial orbits can escape as fast or faster than they are created, leading to tangential anisotropy. As it is harder to escape from the cluster when increasing $\lambda$, more stars on radial orbits will build up, which could explain why our $\lambda$=2 simulation develops radial anisotropy. It was also shown by \citet{Oh1992} that the interaction with the tidal field increases the angular momentum of stars in the outer regions of clusters, causing a reduction in the eccentricity of their orbits. For their simulations this led to a reduction of radial anisotropy towards isotropy; in our case, due to the extended initial conditions, this could lead to an increase in the tangential anisotropy. However it is not known how this would effect would change with $\lambda$ or if it could explain the less tangential anisotropy with increasing $\lambda$.

We then explore the rotation curve of the PEs, by looking at the $\varphi$ component of the velocity in spherical coordinates. Fig.~\ref{fig:vphi_norm} shows the radial profile of $<v_{\varphi}>$ for all the stars (solid lines) and only the PEs (dashed) binned in cylindrical shells in the $xy$ plane, normalised to $\Omega r_{\rm J}$ to see the amount of rotation as a fraction of the total velocity at $r_{\rm J}$. The left-hand and right-hand panels are at the same moments considered in Fig.~\ref{fig:Betapot} and also take the mean of two orbits as explained previously. The PEs have a negative, i.e. retrograde, rotation and as the bound stars have values between 0 and -0.1 $<v_{\varphi}>/\Omega r_{\rm J}$, the rotation of the cluster becomes more negative and retrograde with increasing distance from the centre as PEs increasingly dominate. This negative rotation is expected as retrograde orbits are more stable against escape (\citealt{Keenan1975}; \citealt{Weinberg1991}). The difference between the left-hand and right-hand panels of Fig.~\ref{fig:vphi_norm} shows that over time the fraction of retrograde rotation for the $\lambda$=0 and $\lambda$=1 simulations stays roughly constant at 0.5$\Omega r_{\rm J}$, as seen in \citet*{Tiongco2016b}, but the $\lambda$=2 simulation becomes more negative.

In Section 4, we derive a relation for the velocity dispersion at the Jacobi surface,  $\sigma_{\rm J}$. If we instead normalise $<v_{\varphi}>$ to $\sigma_{\rm J}$, the profile is almost identical to Fig.~\ref{fig:vphi_norm} and can be used to study the relationship between our expression for the velocity dispersion and the rotation in the cluster.

\begin{figure}
    \centering
    \includegraphics[width=1.0\columnwidth]{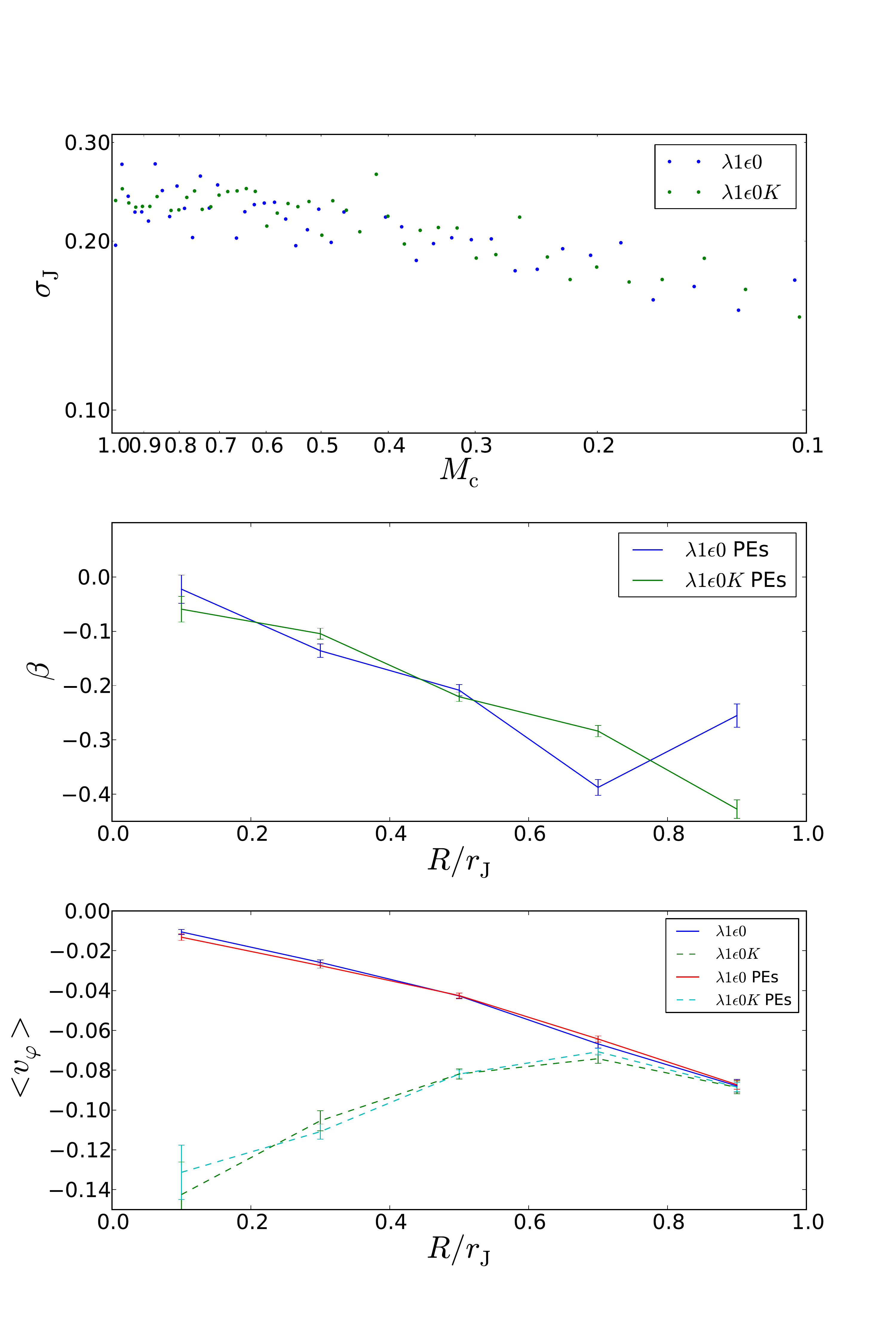}
    \caption{Comparison of the kinematics in the $\lambda$1$\epsilon$0 and $\lambda$1$\epsilon$0K simulations. From top to bottom: dispersion in a spherical bin of $0.9r_{\rm J}$ to $r_{\rm J}$ over the lifetime of the simulations, velocity dispersion anisotropy and $<v_{\varphi}>$ (both at a snapshot when there is 0.5 $M_0$). These quantities have been calculated in the same way as Figs.~\ref{fig:dispprofile},~\ref{fig:Betapot}, and~\ref{fig:vphi_norm}, respectively.}
    \label{fig:imfcomp}
\end{figure}

\subsubsection{IMF dependence}

Fig.~\ref{fig:imfcomp} compares the kinematics of the $\lambda$1$\epsilon$0 and $\lambda$1$\epsilon$0K simulations. In the top panel we show the mass-weighted velocity dispersion in a spherical bin between $0.9r_{\rm J}$ and $r_{\rm J}$, against cluster mass over the lifetime of the simulations. The middle and bottom panels show the $\beta$ and $<v_{\varphi}>$ profiles respectively, at $0.5M_{\rm 0}$ calculated in the same way as Figs.~\ref{fig:Betapot} and~\ref{fig:vphi_norm}. There are minimal differences when changing the mass function, showing that for simulations with the same $<m>$ changing the IMF has no effect on these aspects of the kinematics.

\subsection{Eccentric orbits}

\begin{figure}
    \centering
    \includegraphics[width=1\columnwidth]{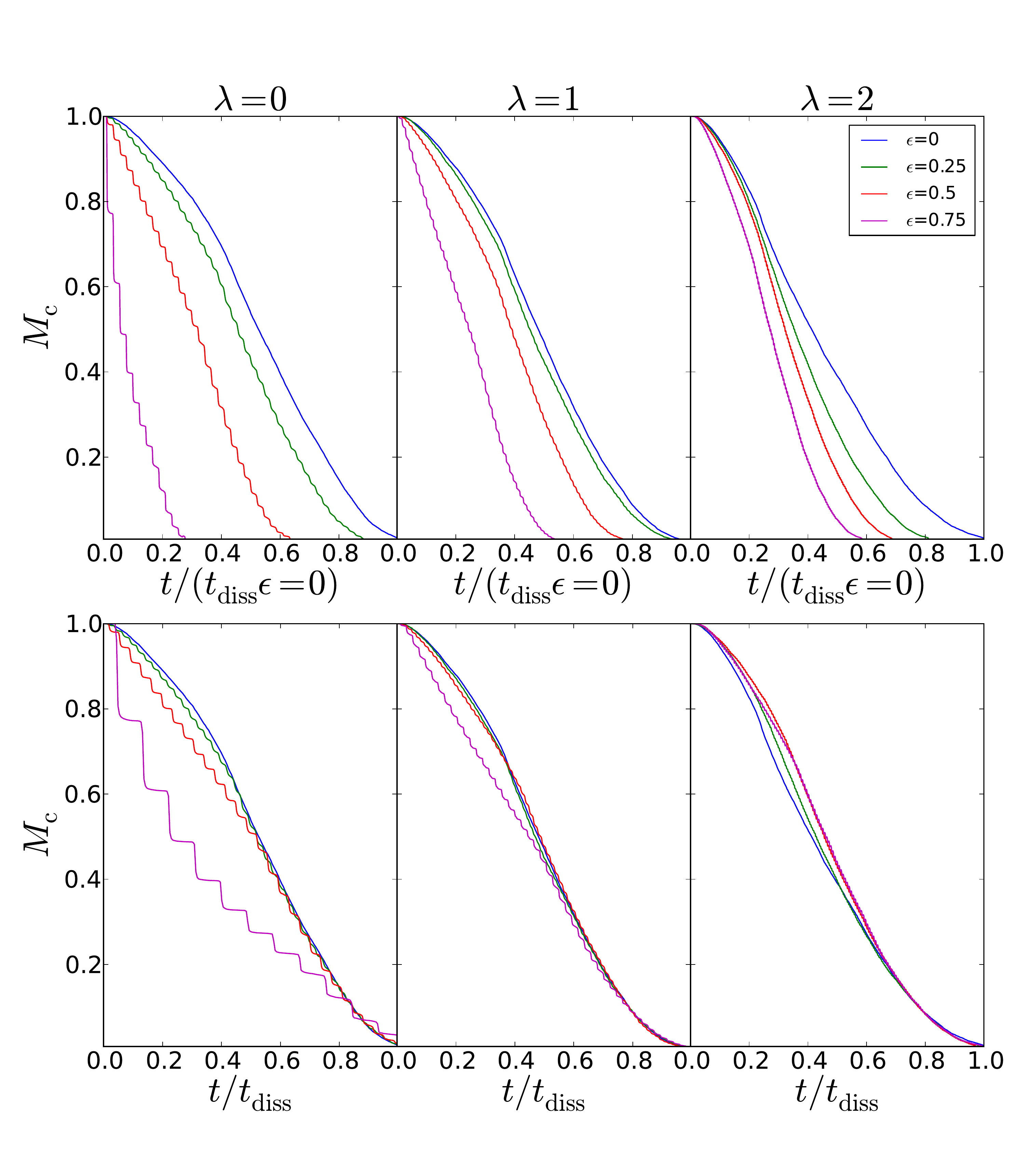}
    \caption{Bound mass evolution for eccentricities of 0, 0.25, 0.5, and 0.75 in each potential. Top plots are from the simulations where we used the same semi-major axis and mean galactocentric distance to approximate the same lifetime. Bottom plots are after scaling the simulations to the same time to reach 0.1 of the initial mass remaining.}
    \label{fig:sis_mbound}
\end{figure}
We now consider the effect of introducing eccentricity to the orbits. Fig.~\ref{fig:sis_mbound} shows the total mass evolution for the equal-mass clusters in all potentials and eccentricities. The top panels show the actual evolution in the different potentials for each eccentricity. All orbits had the same semi-major axis, which ensures that the lifetimes are the same at low $\epsilon$, but for larger eccentricities additional scaling is required to achieve the same lifetimes. \citet{Cai2016} compared $t_{\rm diss}$ of clusters in $\lambda=0$ and $\lambda=1$ galaxies, finding that the eccentricity dependence was smaller for $\lambda=1$. Here we confirm this and find that for $\lambda=2$ the effect of eccentricity is also less important. To achieve the same lifetimes we take a scale-factor of the ratio of the dissolution time of the circular orbit to the eccentric simulation that requires scaling, $T_* = {t_{\rm diss}(\epsilon=0)/{t_{\rm diss}(\epsilon>0)}}$, with $t_{\rm diss}$ taken to be when $M_{\rm c} = 0.1 M_{\rm 0}$ and find the scale parameters for position, velocity and angular velocity as $r_* = T_*^{2/3}$, $v_* = T_*^{-1/3}$ and $\Omega_* =T_*^{-1}$. The bottom panels of Fig.~\ref{fig:sis_mbound} show the scaled mass evolution as a function of scaled time. The early evolution of the $\lambda0\epsilon0.75$ simulation is quite different from the others, and this is likely due to the rapid loss of stars at pericentre. The lower eccentricity orbits match the circular orbit profile later in the lifetime of the simulations.

\begin{figure}
    \centering  
      \includegraphics[width=0.99\columnwidth]{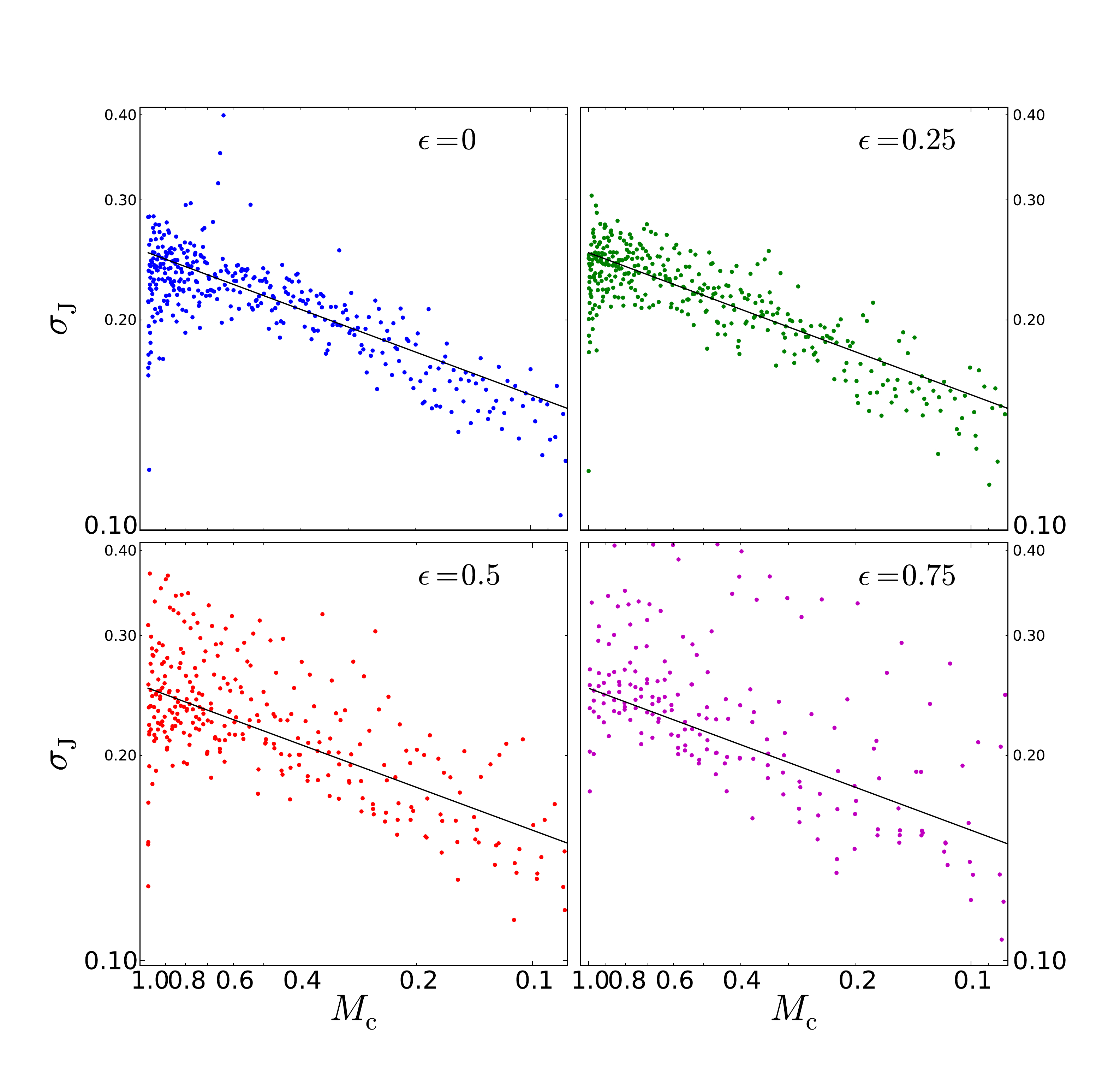}
      \caption{Velocity dispersion in the 0.9$r_{J}$ to $r_{J}$ region in the $\lambda$1$\epsilon$0, $\lambda$1$\epsilon$0.25, $\lambda$1$\epsilon$0.5 and $\lambda$1$\epsilon$0.75 simulations. Black lines are our prediction using the mass and angular velocity of the circular orbit (see Section 4).}
      \label{fig:sis_eccen_scaled}
\end{figure}
\subsubsection{Velocity dispersion and anisotropy}
Fig.~\ref{fig:sis_eccen_scaled} shows the velocity dispersion for stars between $0.9r_{\rm J}$ and $r_{\rm J}$ for the $\lambda$=1 simulations for different $\epsilon$. As the dissolution times of the eccentric orbits have been scaled to be the same as the circular orbit, the $\Omega$ of the circular orbit can be used to approximate that of the eccentric orbits. This gives a smoothly declining $r_{\rm J, circ}$ and mass of the cluster, which we use to calculate our prediction in Section 4 (black lines), and reduces the variation of the dispersion over one orbit. The dispersion is very similar for each simulation, but has an orbital variation that increases with eccentricity. The higher dispersion values are due to a sharp increase at pericentre, but the cluster actually spends most of its time at apocentre and therefore at the lower values of the dispersion. Fig.~\ref{fig:sis_eccen_scaled} shows that the black line prediction well matches the average velocity dispersion of an orbit at any point in the lifetime of the eccentric orbit simulations.
\begin{figure}
    \centering
    \includegraphics[width=0.99\columnwidth]{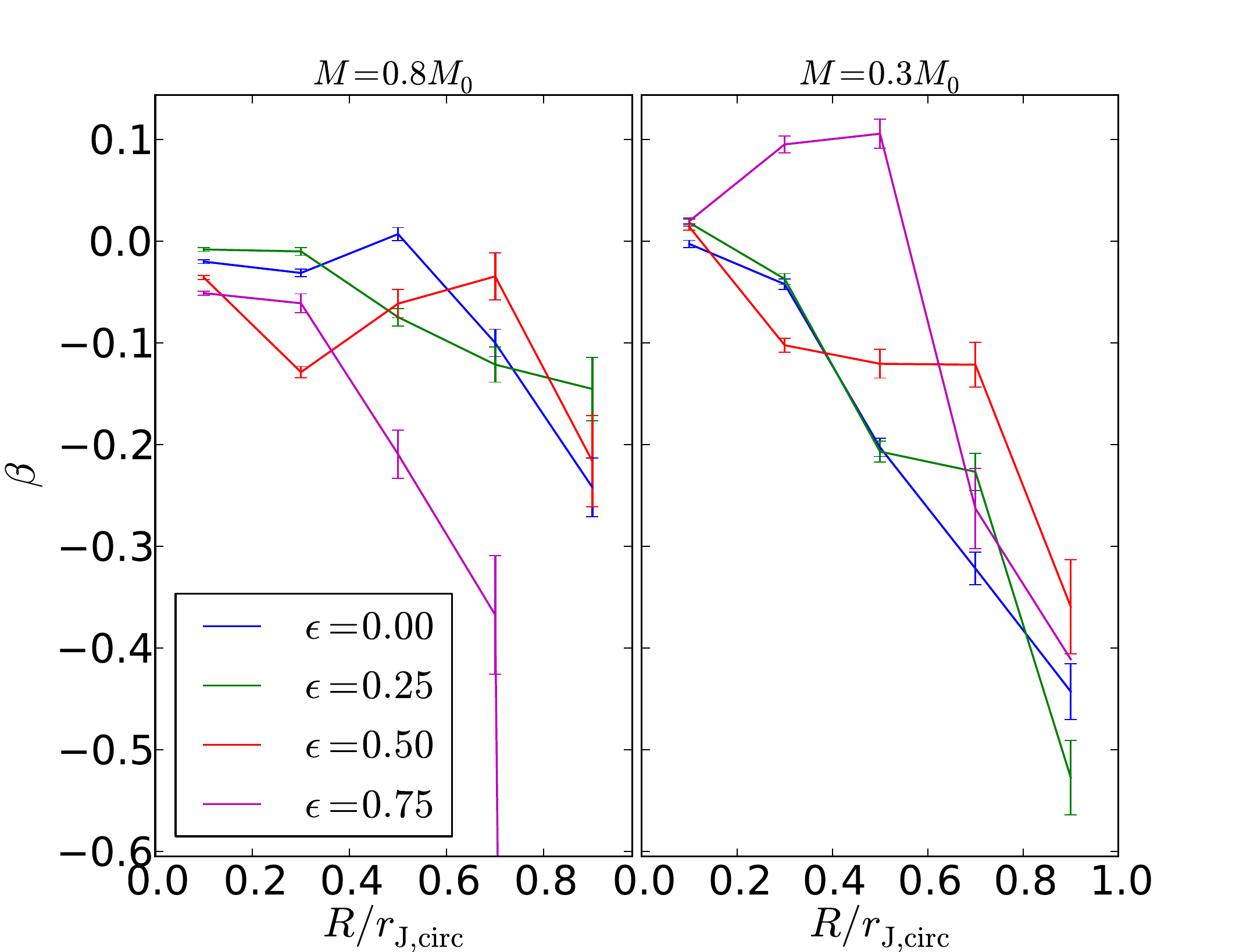}
    \caption{Comparison of the radial profile of the anisotropy for the $\lambda$1$\epsilon$0.25, $\lambda$1$\epsilon$0.5 and $\lambda$1$\epsilon$0.75 eccentric orbit simulations to the circular orbit $\lambda$1$\epsilon$0 simulation, using the mean of snapshots for three orbits around the specified remaining mass for each radial bin.}
    \label{fig:betaeccen}
\end{figure}

Fig.~\ref{fig:betaeccen} shows the $\beta$ profile as a function of $R$ for all stars in the $\lambda$=1 simulations using the approximate value of the Jacobi radius, $r_{\rm J, circ}$. The panels are produced as in Fig.~\ref{fig:Betapot} but only showing the profile for all stars in each bin. As the anisotropy in cylindrical shells is not dependent on the reference frame, and because the majority of the anisotropy is due to the PEs, as shown in Fig.~\ref{fig:Betapot}, this means that variations in the anisotropy profiles across the different eccentricities can be inferred to be variations of the population of PEs, assuming bound stars have an isotropic velocity distribution. Fig.~\ref{fig:betaeccen} shows some variation across the eccentricities for the snapshot later in the lifetime, with less tangential anisotropy when increasing $\epsilon$. The $\epsilon$=0.75 simulation has a very different profile but this is possibly due to the different mass evolution shown in Fig.~\ref{fig:sis_mbound}, as different values for the initial filling factor can lead to variations in the anisotropy as explained earlier.

\subsubsection{Rotation}
\begin{figure}
	\centering
	\includegraphics[width=0.99\columnwidth]{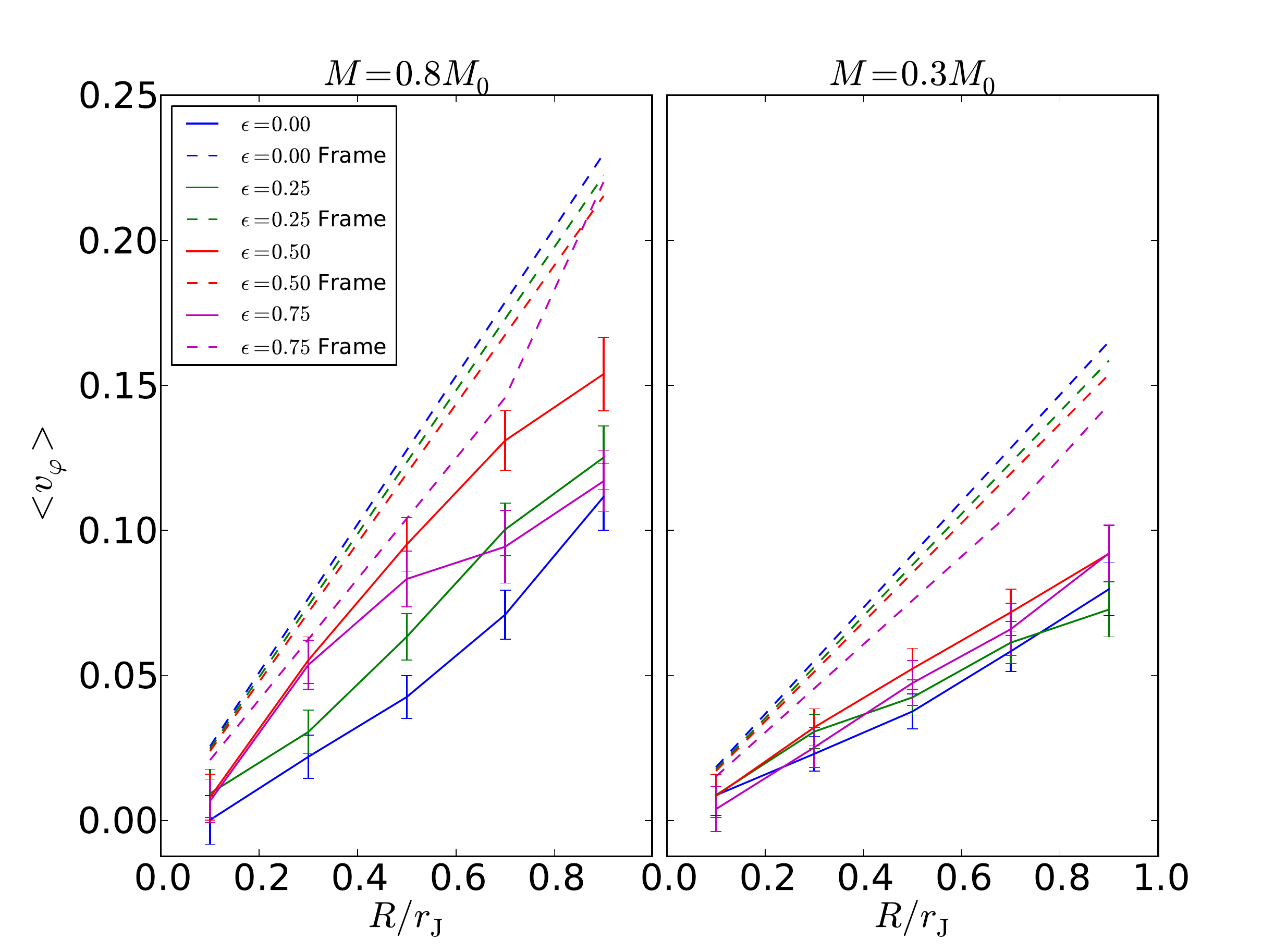}
	\caption{Radial profiles of $<v_{\varphi}>$ for all the stars in the simulations with a $\lambda$=1 potential with different eccentricities. Left-hand panel is the mean of snapshots for three orbits around 0.8 $M_0$. Right-hand panel is the same for 0.3 $M_0$. }
	\label{fig:vphi_eccen}
\end{figure}

For the circular orbits we found that the $<v_{\rm \varphi}>$ of stars near $r_{\rm J}$ is about $0.5\Omega r_{\rm J}$ and retrograde with respect to the orbit. This implies that in a non-rotating frame these stars are on prograde orbits. Fig.~\ref{fig:vphi_eccen} shows the $<v_{\varphi}>$ profile for all stars in the equal-mass $\lambda$=1 case (solid lines) and the mean rotation profile of the frame calculated as $\Omega r$ (dashed lines)\footnote{In the case of the eccentric orbit with $\epsilon =0.75$, the last bin shows a larger rotation than expected from extrapolating the solid-body rotation outwards. This is due to one snapshot not having any stars in that bin and being excluded from the mean. This snapshot corresponded to apocentre where the rotation is at a minimum, and therefore the rotation is higher by not including this snapshot.}. The profiles are calculated again using cylindrical shells in the $xy$ plane. Here however we consider radial positions divided by $r_{\rm J}$ calculated from equation~(\ref{eqn:tidalradius}) for each snapshot. We chose this normalisation because the features of the rotation are washed out when using $r_{\rm J, circ}$  as the cluster expands and contracts over the course of an orbit.

From Fig. 13 we see that the $<v_{\varphi}>$ profiles are similar for different $\epsilon$, which at $r_{\rm J}$ are close to the $0.5\Omega r_{\rm J}$ found in Fig.~\ref{fig:vphi_norm}. The left-hand panel shows that early in the simulation there is some variation with eccentricity, as the eccentric orbits have higher $<v_{\varphi}>$ than the circular orbit, but this variation seems to decrease with time. The solid-body rotation of the frame, $\Omega r$ (dashed lines), also varies as it decreases with increasing eccentricity. This means if we subtract the solid-body rotation of the frame from the $<v_{\phi}>$ of the stars, to convert to a fiducial reference frame that rotates at $\Omega_{\rm circ}$, there would be less retrograde rotation in clusters on higher $\epsilon$ orbits.

\section{Velocity dispersion at the Jacobi surface}

\subsection{Derivation}
B01 derived a relation for $N(\hat{E})$ (equation~\ref{eqn:Ne}). We can use this result to derive a relation for the velocity dispersion of the PEs. As $N(\hat{E})$ is a probability density function, the mean can be found from
\begin{equation}
	\left\langle\hat{E}\right\rangle = \int_{0}^{\infty}\hat{E}N(\hat{E})d\hat{E} \propto  \left(\frac{t_{\rm esc}}{t_{\rm rh}}\right)^{1/4},
	\label{eq:meanEhat}
\end{equation}
including our extra $\lambda$ dependence from Section 3.2.3. If we relate the energy to velocity using $\hat{E} \propto v^2/|{E_{\rm crit}}|$ with $E_{\rm J} = (v^2/2) + E_{\rm crit}$ at the Jacobi surface, and assume that the velocity dispersion is related to $<v^2>$ as for a Maxwellian distribution, we can find
\begin{equation}
 \sigma_{\rm J} \propto \sqrt{<v^2>} \propto (<\hat{E}> E_{\rm crit})^{1/2}.
 \label{eq:sig} 
\end{equation}
By substituting equation~(\ref{eq:meanEhat}) into equation~(\ref{eq:sig}) and by using $t_{\rm esc}$ and $t_{\rm rh}$ as defined in Section 3.2.3, and $|E_{\rm crit}| \propto M_{\rm c}/r_{\rm J} \propto (3-\lambda)^{1/3}\Omega^{2/3}M_{\rm c}^{2/3}$, we find

\begin{equation}
\sigma_{\rm J} \propto (3-\lambda)^{-1/12}{M_{\rm c}^{5/24}}\Omega^{1/3}{(<m>\ln{\Lambda})^{1/8}} \left(\frac{r_{\rm hm}}{r_{\rm J}}\right)^{-3/16}.
\label{eq:prediction}
\end{equation}

This can be compared to the MOND prediction which has a $M_{\rm c}^{1/4}$ dependence, very close to the one obtained here. However, equation~(\ref{eq:prediction}) has further dependencies which provide a way of discriminating between the two predictions using observational data.

\begin{figure}
	\centering  
  	\includegraphics[width=0.99\columnwidth]{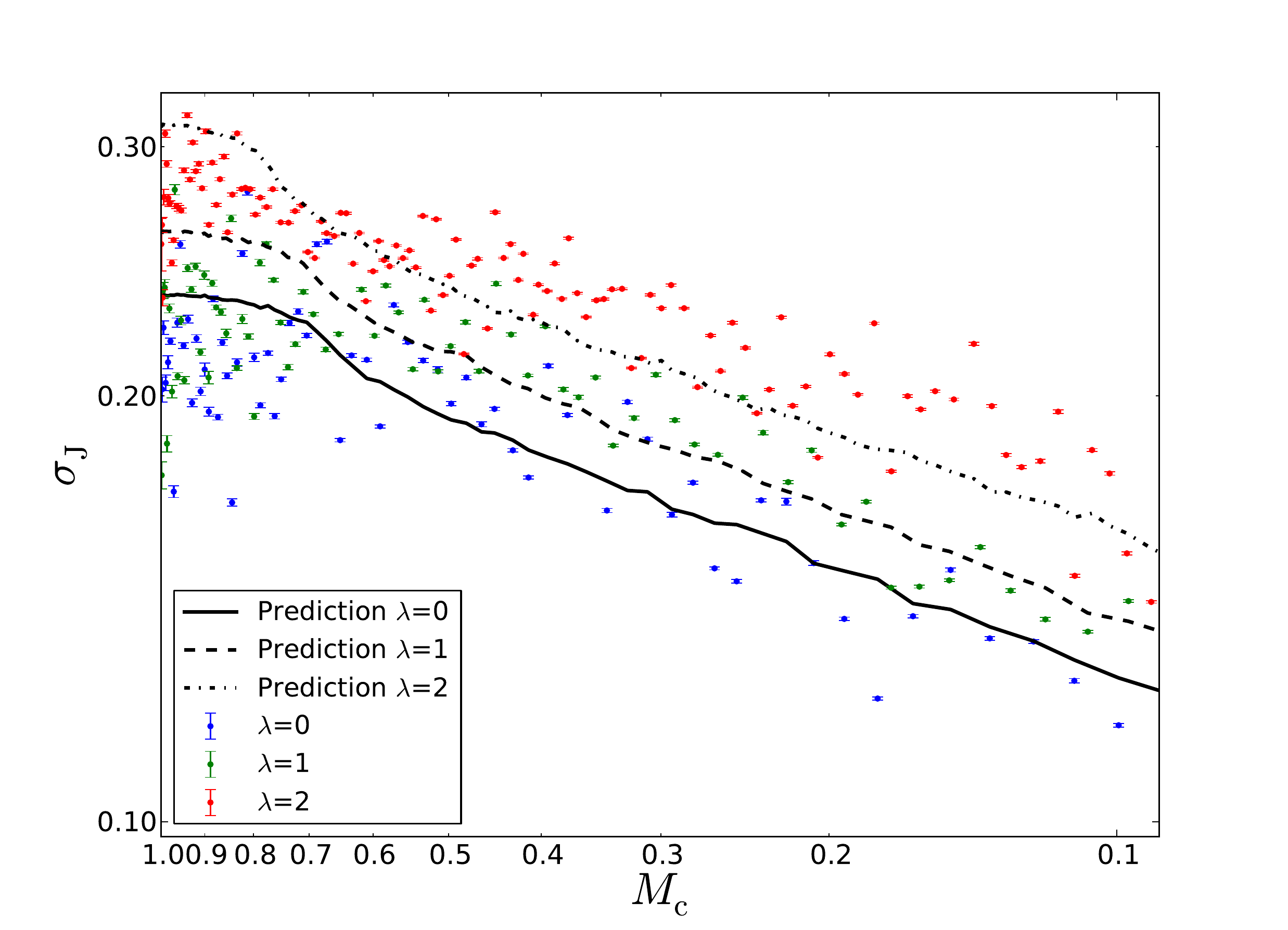}
  	\caption{Velocity dispersion of stars between 0.9$r_{\rm J}$ and $r_{\rm J}$ against remaining mass of the cluster, $M_{\rm c}$, for the $\lambda$0$\epsilon$0, $\lambda$1$\epsilon$0 and $\lambda$2$\epsilon$0 simulations (coloured points). Black lines are the prediction from equation~(\ref{eq:prediction}), with the constant of proportionality fit to the $\lambda$=0 case.}
  	\label{fig:circdisp}
\end{figure}

\begin{figure}
	\centering
	\includegraphics[width=0.99\columnwidth]{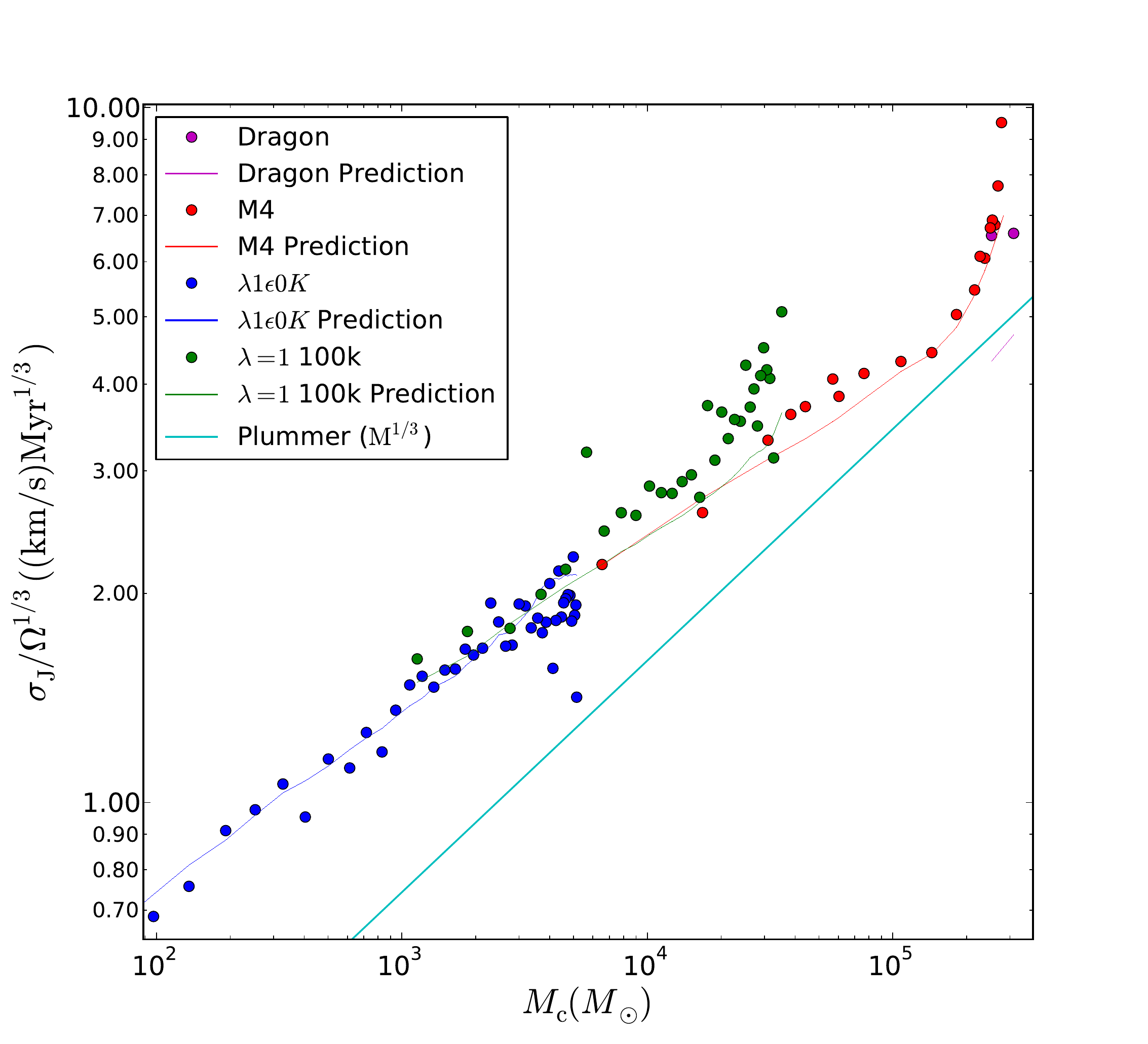}
	\caption{Comparison of our $\sigma_{\rm J}$ prediction to simulations with larger number of particles. The dispersion has been divided by $\Omega^{1/3}$ for each simulation to reduce the largest difference so the profiles can more easily be compared.}
	\label{fig:disp_largeN}
\end{figure}

\subsection{Comparison of the velocity dispersion prediction to simulations}

To establish whether our derived scaling of $\sigma_{\rm J}$ in equation~(\ref{eq:prediction}) holds in our $N$-body simulations, we compare our prediction to the dispersion of the stars near $r_{\rm J}$ for the circular orbits in each potential. We focus on a spherical shell between $0.9r_{\rm J}$ and $r_{\rm J}$ as there will only be PEs in this region of the cluster.

Fig.~\ref{fig:circdisp} shows the velocity dispersion of stars in this shell as a function of the mass of the cluster for the $\lambda$0$\epsilon$0, $\lambda$1$\epsilon$0 and $\lambda$2$\epsilon$0 simulations. The black lines reproduce our predictions from equation~(\ref{eq:prediction}), after finding the constant of proportionality by fitting to the $\lambda$=0 case. The velocity dispersion near $r_{\rm J}$ from the simulations increases with higher $\lambda$ which is well reproduced by our $\lambda$ dependence. The mass dependence also accurately reproduces the decline in $\sigma_{\rm J}$ as $M_{\rm c}$ decreases. 

There is a large amount of scatter in the values from the simulation that could accommodate a range of mass dependencies. We therefore also compare our prediction to simulations with higher number of particles. Fig~\ref{fig:disp_largeN} shows the dispersion, divided by $\Omega^{1/3}$ to remove the largest variation between simulations, against the remaining mass of the cluster over time for the $\lambda$1$\epsilon$0K simulation (blue), an $N_{0} = 10^{5}$ particle simulation run for the Gaia Challenge Workshop (green, http://bit.ly/241CBMJ, \citealt{Peuten2016}), a simulation of the cluster M4 (red, \citealt{Heggie2014}) and the $N=10^6$ particle Dragon simulations (magenta, \citealt{Wang2016}). The predictions for each simulation are plotted using the constant of the fit from Fig.~\ref{fig:circdisp} (solid lines). Our prediction slightly underestimates the 100k and Dragon simulations, but matches the M4 simulation extremely well, including the compact initial conditions and the subsequent expansion to fill the Roche volume. The difference between the predictions for each simulation shows the importance of including the $<m>$ dependence in our prediction. Even though the M4 simulation has a $\lambda$=0 galactic potential and should therefore have a lower $\sigma_{J}$, it has a higher mean mass which increases $\sigma_{\rm J}$. This can also be seen in the difference between the prediction for the M4 simulation and Dragon simulation, as despite the latter also using a point mass potential it has a lower $<m>$ and much more extended intial filling factor. The discrepency between the value of $\sigma_{\rm J}$ and the prediction for the Dragon simulation is possibly due to the cluster having an initial population of remnants that are dynamically unevolved in these snapshots.

We also over-plot the prediction of the velocity dispersion for a Plummer model (magenta line), $\sigma = \sqrt{GM_{\rm c}/(6\sqrt{r^2 +r_0^2})}$, at $r_{\rm J}$ and using $r_0 \sim r_{\rm hm}/1.3$ (see page 73 of \citealt{Heggie2003}): 
\begin{equation}
	\sigma_{\rm J} = \frac{2^{1/6}}{6^{1/2}}(GM_c\Omega)^{1/3} \left[1+\left(\frac{r_{\rm hm}}{r_{\rm J}}\right)^2 \right]^{-1/4}
\end{equation}
and adopting $r_{\rm hm}/r_{\rm J}$ as $\sim 0.15$. This also has an $\Omega^{1/3}$ dependence like our prediction, and underpredicts the dispersion for most masses. Due to a steeper $M_{\rm c}^{1/3}$ dependence, this relation approaches $\sigma_{\rm J}$ of equation~(\ref{eq:prediction}) in the mass range of globular clusters ($M_{\rm c} \gtrsim 10^5 {\rm M}_\odot$).

\subsection{Comparison of the velocity dispersion prediction to observational data}

\begin{table*}
\begin{center}
	\caption{Properties of the sample of Milky Way GCs (Column 1 and 2 are from Baumgardt 2017). Columns indicate: name of the cluster, velocity dispersion in outermost bin $\sigma_{\mathrm{lb}}$, radial position of outermost bin $r_{\mathrm{lb}}$, mass of the cluster $M_{\rm c}$, galactocentric radius of the orbit of the cluster $R_{\rm g}$, half-mass radius $r_{\rm hm}$, Jacobi radius $r_{\rm J}$, prediction of the velocity dispersion at the Jacobi surface, $\sigma_{\mathrm{J}}$, ratio of the position of the last bin to Jacobi radius $r_{\mathrm{lb}}/r_{\rm J}$ and ratio of the dispersion in the last bin to the prediction of the dispersion $\sigma_{\mathrm{lb}}/\sigma_{\mathrm{J}}$.}
	\begin{tabular}{|l|r|r|r|r|r|r|r|r||r|r}

	\hline
	Cluster & $\sigma_{\mathrm{lb}}$  & $r_{\mathrm{lb}}$ &  $M_{\rm c}$ & $R_{\rm g}$ & $r_{\rm hm}$ & $r_{\rm J}$ & $\sigma_{\mathrm{J}}$ & $r_{\mathrm{lb}}/r_{\rm J}$ & $\sigma_{\mathrm{lb}}/\sigma_{\mathrm{J}}$\\
	& $\mathrm{km\,s^{-1}}$ & pc & $10^5M_{\rm \odot}$ & kpc &pc&pc  & km$\,s^{-1}$ &  & \\ \hline
NGC104 & $4.58^{+0.42}_{-0.36}$ & 54.27 & 10.02 & 7.40 & 6.82 & 117.580 & 1.55 & 0.46 & $2.96^{+0.27}_{-0.23}$\\
NGC288 & $1.77^{+0.20}_{-0.18}$ & 21.80 & 0.86 & 12.00 & 7.78 & 71.499 & 0.68 & 0.30 & $2.60^{+0.29}_{-0.26}$\\
NGC362 & $2.93^{+0.69}_{-0.51}$ & 17.63 & 4.00 & 9.40 & 2.24 & 101.764 & 1.40 & 0.17 & $2.09^{+0.61}_{-0.45}$\\
NGC1851 & $3.11^{+0.56}_{-0.44}$ & 35.38 & 3.67 & 16.60 & 2.46 & 144.184 & 1.19 & 0.25 & $2.61^{+0.38}_{-0.30}$\\
NGC1904 & $2.12^{+0.30}_{-0.25}$ & 33.85 & 2.38 & 18.80 & 3.55 & 135.607 & 0.96 & 0.25 & $2.21^{+0.31}_{-0.26}$\\
NGC2419 & $1.30^{+1.01}_{-3.62}$ & 160.08 & 10.02 & 89.90 & 23.27 & 621.370 & 0.73 & 0.26 & $1.78^{+1.38}_{-4.95}$\\
NGC2808 & $5.61^{+0.69}_{-0.57}$ & 23.12 & 9.75 & 11.10 & 2.58 & 152.660 & 1.69 & 0.15 & $3.31^{+0.41}_{-0.34}$\\
NGC3201 & $2.31^{+0.27}_{-0.23}$ & 38.43 & 1.63 & 8.80 & 7.94 & 72.083 & 0.87 & 0.53 & $2.66^{+0.31}_{-0.26}$\\
NGC4147 & $1.62^{+0.41}_{-0.30}$ & 19.28 & 0.50 & 21.40 & 2.99 & 87.994 & 0.62 & 0.22 & $2.61^{+0.66}_{-0.48}$\\
NGC4372 & $3.21^{+0.40}_{-0.33}$ & 14.61 & 2.23 & 7.10 & 8.08 & 69.345 & 0.99 & 0.21 & $3.24^{+0.40}_{-0.33}$\\
NGC4590 & $0.74^{+0.52}_{-0.40}$ & 25.63 & 1.52 & 10.20 & 4.48 & 77.609 & 0.92 & 0.33 & $0.81^{+0.57}_{-0.44}$\\
NGC4833 & $3.48^{+0.46}_{-0.38}$ & 8.03 & 3.17 & 7.00 & 4.91 & 77.193 & 1.20 & 0.10 & $2.89^{+0.38}_{-0.32}$\\
NGC5024 & $2.05^{+0.42}_{-0.32}$ & 72.49 & 5.21 & 18.40 & 7.01 & 173.536 & 1.06 & 0.42 & $1.94^{+0.40}_{-0.30}$\\
NGC5053 & $1.02^{+0.25}_{-0.20}$ & 40.37 & 0.87 & 17.80 & 13.51 & 93.280 & 0.57 & 0.43 & $1.80^{+0.44}_{-0.35}$\\
NGC5139 & $7.60^{+0.37}_{-0.34}$ & 58.00 & 21.73 & 6.40 & 9.31 & 138.133 & 1.87 & 0.42 & $4.06^{+0.20}_{-0.18}$\\
NGC5272 & $2.43^{+0.48}_{-0.36}$ & 54.48 & 6.10 & 12.00 & 8.06 & 137.499 & 1.18 & 0.40 & $2.06^{+0.41}_{-0.31}$\\
NGC5286 & $7.45^{+0.85}_{-0.71}$ & 6.06 & 5.36 & 8.90 & 1.89 & 107.921 & 1.59 & 0.06 & $4.69^{+0.53}_{-0.44}$\\
NGC5466 & $0.99^{+0.26}_{-0.20}$ & 50.67 & 1.06 & 16.30 & 10.91 & 94.110 & 0.64 & 0.54 & $1.56^{+0.41}_{-0.31}$\\
NGC5694 & $2.57^{+0.50}_{-0.39}$ & 33.05 & 2.32 & 29.40 & 3.42 & 181.027 & 0.87 & 0.18 & $2.94^{+0.57}_{-0.45}$\\
NGC5824 & $3.70^{+0.77}_{-0.59}$ & 30.11 & 5.93 & 25.90 & 3.39 & 227.533 & 1.17 & 0.13 & $3.16^{+0.66}_{-0.50}$\\
NGC5904 & $2.91^{+0.37}_{-0.31}$ & 26.31 & 5.72 & 6.20 & 3.19 & 86.651 & 1.58 & 0.30 & $1.84^{+0.24}_{-0.20}$\\
NGC5927 & $4.09^{+0.60}_{-0.49}$ & 8.27 & 2.28 & 4.60 & 1.47 & 52.242 & 1.50 & 0.16 & $2.73^{+0.40}_{-0.33}$\\
NGC6093 & $6.38^{+0.43}_{-0.39}$ & 1.99 & 3.35 & 3.80 & 0.67 & 52.324 & 2.01 & 0.04 & $3.17^{+0.21}_{-0.19}$\\
NGC6121 & $3.30^{+0.24}_{-0.22}$ & 17.98 & 1.29 & 5.90 & 7.43 & 50.981 & 0.89 & 0.35 & $3.70^{+0.27}_{-0.25}$\\
NGC6139 & $6.43^{+1.23}_{-0.96}$ & 3.57 & 3.78 & 3.60 & 0.89 & 52.527 & 2.00 & 0.07 & $3.22^{+0.62}_{-0.48}$\\
NGC6171 & $2.42^{+0.33}_{-0.28}$ & 8.47 & 1.21 & 3.30 & 1.66 & 33.873 & 1.31 & 0.25 & $1.85^{+0.25}_{-0.21}$\\
NGC6205 & $4.01^{+0.46}_{-0.39}$ & 24.97 & 4.50 & 8.40 & 4.13 & 97.956 & 1.32 & 0.25 & $3.04^{+0.35}_{-0.30}$\\
NGC6218 & $2.67^{+0.46}_{-0.37}$ & 74.17 & 1.44 & 29.80 & 15.34 & 155.712 & 0.57 & 0.48 & $4.65^{+0.80}_{-0.64}$\\
NGC6254 & $2.95^{+0.58}_{-0.46}$ & 9.15 & 1.68 & 4.50 & 2.55 & 46.521 & 1.25 & 0.20 & $2.37^{+0.20}_{-0.37}$\\
NGC6273 & $9.13^{+1.48}_{-1.19}$ & 3.41 & 7.67 & 4.60 & 1.77 & 78.346 & 2.04 & 0.04 & $4.47^{+0.72}_{-0.58}$\\
NGC6341 & $3.19^{+0.39}_{-0.33}$ & 4.24 & 3.29 & 1.70 & 0.50 & 30.419 & 2.50 & 0.14 & $1.28^{+0.16}_{-0.13}$\\
NGC6388 & $7.28^{+0.93}_{-0.77}$ & 16.59 & 9.93 & 9.60 & 1.45 & 139.431 & 1.95 & 0.12 & $3.73^{+0.48}_{-0.39}$\\
NGC6397 & $3.20^{+0.21}_{-0.19}$ & 8.18 & 0.77 & 3.10 & 2.62 & 28.038 & 1.08 & 0.29 & $2.98^{+0.20}_{-0.18}$\\
NGC6402 & $6.08^{+0.92}_{-0.24}$ & 10.15 & 7.47 & 6.00 & 2.27 & 92.671 & 1.83 & 0.11 & $3.33^{+0.50}_{-0.40}$\\
NGC6656 & $3.36^{+0.59}_{-0.54}$ & 17.68 & 4.30 & 4.00 & 3.91 & 58.823 & 1.54 & 0.30 & $2.19^{+0.38}_{-0.35}$\\
NGC6715 & $8.80^{+1.50}_{-1.50}$ & 21.66 & 16.79 & 4.90 & 1.17 & 106.082 & 2.71 & 0.20 & $3.24^{+0.55}_{-0.55}$\\
NGC6723 & $2.83^{+0.31}_{-0.50}$ & 20.92 & 2.32 & 18.90 & 8.41 & 134.840 & 0.81 & 0.16 & $3.50^{+0.79}_{-0.62}$\\
NGC6752 & $2.76^{+0.30}_{-0.28}$ & 9.09 & 2.11 & 2.60 & 1.44 & 34.847 & 1.66 & 0.26 & $1.66^{+0.19}_{-0.17}$\\
NGC6809 & $3.70^{+0.34}_{-0.27}$ & 11.76 & 1.82 & 5.20 & 4.28 & 52.664 & 1.12 & 0.22 & $3.29^{+0.27}_{-0.24}$\\
NGC6838 & $1.02^{+0.39}_{-0.23}$ & 6.58 & 0.30 & 3.90 & 1.89 & 23.817 & 0.83 & 0.28 & $1.23^{+0.41}_{-0.28}$\\
NGC7078 & $3.03^{+0.21}_{-0.19}$ & 14.71 & 8.11 & 6.70 & 1.95 & 102.540 & 1.88 & 0.14 & $1.61^{+0.11}_{-0.10}$\\
NGC7089 & $3.92^{+0.64}_{-0.51}$ & 26.27 & 7.00& 10.40 & 3.21 & 130.878 & 1.50 & 0.20 & $2.61^{+0.43}_{-0.34}$\\
NGC7099 & $2.12^{+0.25}_{-0.22}$ & 14.96 & 1.63 & 7.10 & 2.13 & 62.472 & 1.16 & 0.24 & $1.82^{+0.21}_{-0.19}$\\
Ter8 & $1.46^{+0.47}_{-0.40}$ & 17.35 & 0.18 & 19.40 & 5.36 & 58.800 & 0.42 & 0.30 & $3.44^{+1.11}_{-0.94}$\\ \hline
	\end{tabular}
\end{center}
\end{table*}

\begin{figure}
	\centering
	\includegraphics[width=0.99\columnwidth]{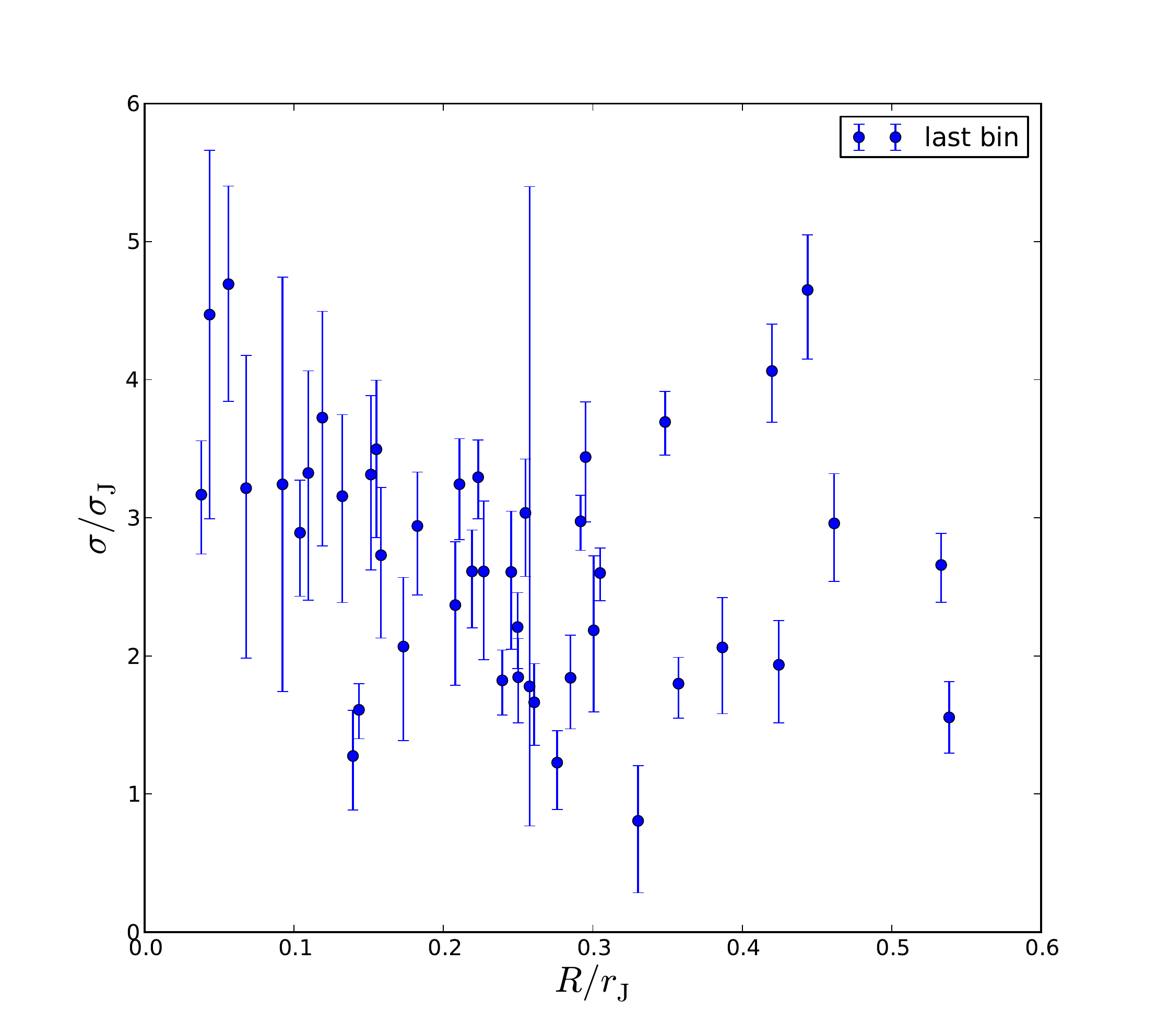}
	\caption{Ratio of the velocity dispersion in the last bin of the profiles from the Baumgardt (2016) data, to our prediction $\sigma_{J}$, as a function of the ratio of the position of the last bin to the calculated $r_{\rm J}$ (blue). The bins with the lowest values of velocity dispersion found in each of the profiles are also plotted in green.}
	\label{fig:obsfract}
	
\end{figure}
\begin{figure}
	\centering
	\includegraphics[width=0.99\columnwidth]{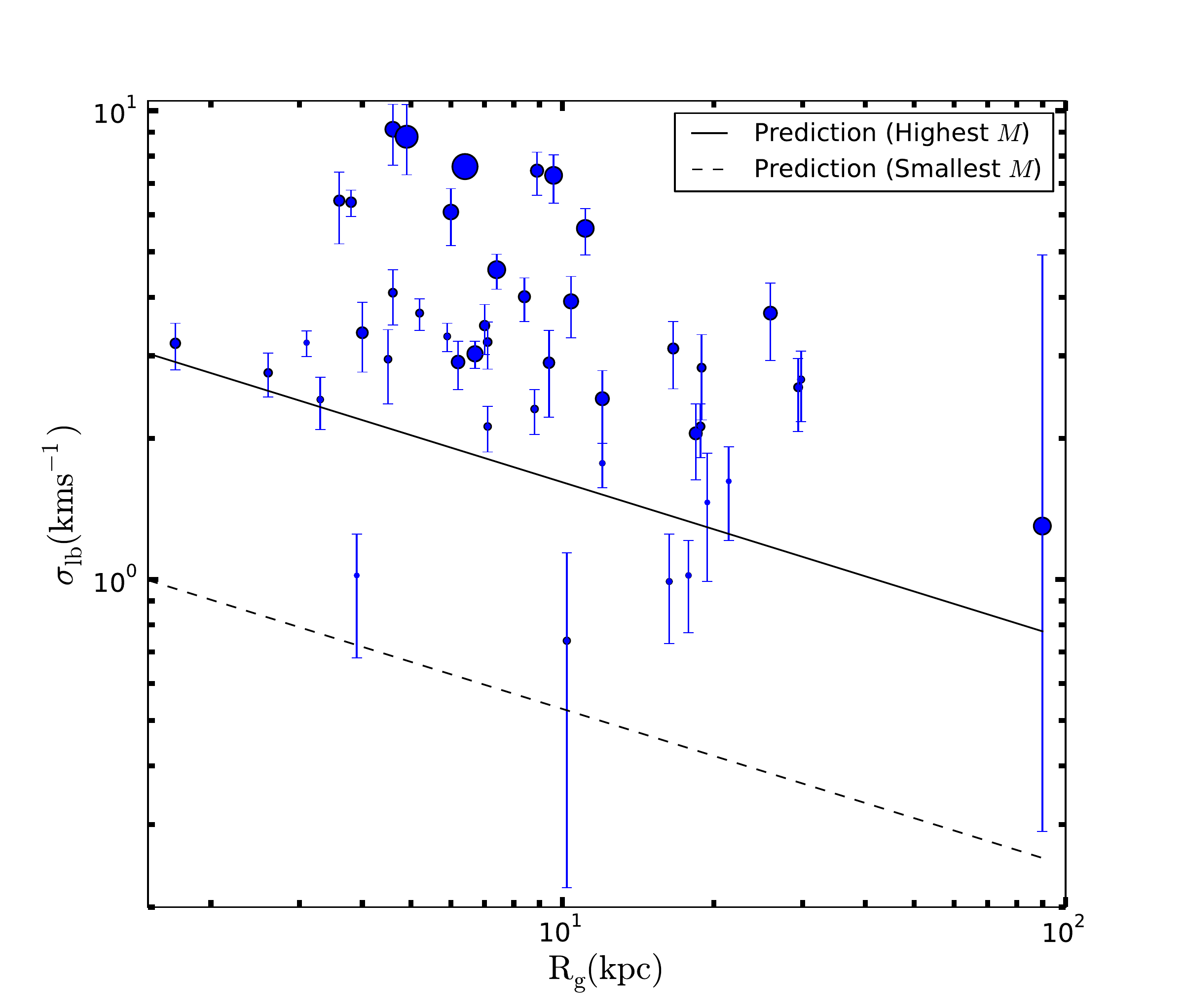}
	\caption{Velocity dispersion of the last bin of data for the clusters from Baumgardt (2016), plotted against the galactocentric distance of the cluster). The size of the points is proportional to the mass of the clusters. The black lines are our prediction for the most massive and least massive clusters (solid and dashed respectively).}
	\label{fig:Rdep}
\end{figure}
It is also possible to directly compare our prediction of $\sigma_{\rm J}$ to observational data. \citet{Baumgardt2017} presented a compilation of line-of-sight velocities and proper motion data for stars in 50 Milky Way GCs from a wide range of data available from literature, which was used to create combined velocity dispersion profiles.

We can consider the outermost bin of each of these velocity dispersion profiles and compare them to the value obtained from our prediction for each cluster. To calculate our estimate of $\sigma_{\rm J}$ we approximate the mass of the cluster using the absolute visual magnitude from the Harris catalogue (\citealt{Harris1996}; 2010 edition) and mass-to-light ratio from \citet{McLaughlin2005}. We also estimate the angular velocity of the clusters using $\Omega = V_{\rm c}/R_{\rm g}$, by assuming $V_{\rm c}=220$ km/s and by taking $R_{\rm g}$ from the Harris catalogue.  Table 2 includes the dispersion, $\sigma_{\rm lb}$, and radial position, $R_{\rm lb}$, of the last data point, the ratio of the position of the last bin to the Jacobi radius, $R_{\rm{lb}}/r_{\rm J}$, and ratio of the dispersion in the last bin to the prediction of the dispersion $\sigma_{\mathrm{lb}}/\sigma_{\mathrm{J}}$. 

Figure~\ref{fig:obsfract} shows $\sigma_{\mathrm{lb}}/\sigma_{\mathrm{J}}$ against $R_{\mathrm{lb}}/r_{\rm J}$. It is clear that the data does not extend to the approximate $r_{\rm J}$, which means that the observed dispersions are expected to be higher than the prediction for $\sigma_{\rm J}$. Most points follow an expected trend, with $\sigma_{\mathrm{lb}}/\sigma_{\mathrm{J}}$ decreasing with increasing $R_{\mathrm{lb}}/r_{\mathrm{J}}$. The are some points that appear to not follow this trend. This is possibly due to internal properties of the cluster, which affect the radial distance from the centre of the cluster at which the effects of PEs or of contamination from field or extra-tidal stars become significant. Moreover, we recall that the calculations of $\sigma_{\rm J}$ and $r_{\rm J}$ include some approximations that for some clusters could be less accurate than others.


Fig.~\ref{fig:Rdep} shows the velocity dispersion in the last bin of data against the galactocentric distance for each cluster in our sample, with the point size reflecting the mass of the cluster. We also show our prediction from equation~(\ref{eq:prediction}) for the least massive cluster of the sample (dashed line) and for the most massive (solid line). Although most of the points lie above our prediction, there seems to be an increase in the dispersion with decreasing $R_{\rm g}$, suggesting there is more than just a mass dependence in the velocity dispersion at the Jacobi surface. 
	Our prediction is a lower limit for the dispersion profiles, so it is expected that none of the data points should be lower than our prediction. This is because there are many reasons why the velocity dispersion of the outermost bins in the profile can be above our prediction, including projection effects and observational profiles not extending out to the Jacobi surface. This means that it is likely difficult to discern between the effects that a DM halo may have on the outer regions of a velocity dispersion profile, from the effects of PEs. The proper motion data provided by future releases of the \textit{Gaia} mission will allow for more rigorous selection criteria for cluster members, which can be followed up by further ground based observation of line-of-sight velocities, making it possible to probe closer to the Jacobi surface of Milky Way star clusters. A combination of proper motions and radial velocity measurements for stars in the outer regions of GCs will also provide a way of inspecting the rotation and anisotropy of the dispersion, which may be required to discriminate between these scenarios as the retrograde bias in the orbits of the PEs may not be present when there is the additional effect of a dark matter halo.
	
\section{Conclusions}

By running simulations of star clusters and varying the orbital eccentricities, initial mass function, and galactic (power-law) mass profiles, we have explored the distribution and behaviour of a population of energetically unbound stars within the Jacobi radius of a cluster, and found three properties of the PEs to vary with the slope of the enclosed galactic mass: 1) the fraction of PEs inside the Jacobi radius, 2) the velocity dispersion at the Jacobi surface, 3) the velocity anisotropy.

For an equal-mass system in a point mass galactic potential we found the fraction of PEs inside the Jacobi radius to be consistent with the value found by B01. However, a mass spectrum and shallower galactic density profiles both cause an increase in the number of PEs, up to 40\% in a $1/R_{\rm g}$ density profile galaxy with a globular cluster type IMF between 0.1 and 1$M_{\odot}$. At $r=0.5r_{\rm J}$ there are equal number of PEs and bound stars, and beyond this radius the PEs dominate. This suggests that PEs should have a large influence on cluster kinematics, especially in the outer parts. By inspecting the fraction of total mass in PEs and the evolution of the distribution of masses for the PEs, we found that a large fraction of PEs will be low mass for most of the lifetime of the cluster, meaning that the majority of these stars could not be observed currently, but can contribute significantly to the total mass. 

The energy distribution of PEs becomes wider as $N$ decreases. This width is also larger for larger $\lambda$, and we introduce a $\lambda$ dependence to the model established in B01.

We then investigated the effect of the PEs on the kinematics. The radial profiles of the  anisotropy of the dispersion early in the simulations for the circular orbits are consistent with zero (i.e. isotropy). However, the simulations in the $\lambda$=0 and $\lambda$=1 potentials develop tangential anisotropy in time whereas the $\lambda$=2 simulation shows radial anisotropy. This is possibly due to two-body interactions scattering stars outwards on radial orbits, as these orbits also preferentially escape from the cluster. Therefore the clusters with the larger escape time in the shallowest galactic density profiles create radial orbits faster than the stars can escape. Throughout the entire lifetime the clusters in the $\lambda$=1 and $\lambda$=2 simulations also have some radial anisotropy before $\beta$ decreases towards the tangential anisotropy. This decrease in $\beta$ occurs faster in the $\lambda$=0 simulation. 

The rotation profiles show a clear negative value for the mean of the $\varphi$ component of the velocity in the corotating reference frame which is also seen in a negative bias of the $J_{z}$ distribution. This retrograde motion is expected as prograde orbits are less stable and preferentially lost from the cluster. The PEs cause the $<v_\varphi>$ profile to become increasingly negative with radius, as PEs dominate further from the centre of the cluster, and at the Jacobi radius they have around half of the circular velocity at $r_{\rm J}$. There is also a difference in the $\lambda$=2 simulation, which seems to develop more negative $<v_\varphi>$ over the lifetime, whereas the $\lambda$=0 and $\lambda$=1 stay roughly constant. For the simulations of clusters with a mass spectrum there seems to be no substantial variation in the dynamics when comparing to the equal-mass simulations. Similarly when using higher values of orbital eccentricity, there seems to be only minimal variation of the dynamics, but there is a suggestion of less tangential anisotropy and less retrograde rotation when increasing eccentricity.

We then formulated a relation for the velocity dispersion at $r_{\rm J}$ due to the effect of PEs. From the model of the distribution of $\hat{E}$ of PEs developed in B01, we can approximate the velocity dispersion at the Jacobi surface $\sigma_{\rm J}$ as a function depending on $(3-\lambda)^{-1/12}M_{\rm c}^{5/24}\Omega^{1/3}(<m>\ln{N})^{1/8}$. We compared our prediction to simulations and observational data. By scaling the constant of proportionality of the prediction to match the velocity dispersion between 0.9$r_{\rm J}$ and $r_{\rm J}$ over time in the $\lambda$0$\epsilon$0 simulation, the profile is well matched by the mass dependence of our prediction and the $\lambda$ dependence reproduces the variation across the different potentials. We also found our prediction to be close to the values of the velocity dispersion near the Jacobi radius in simulations with a much larger number of particles. 

This prediction is useful for testing the different theories that attempt to explain the flattening of the velocity dispersion. For example, some predictions using MOND find the flattened value of the velocity dispersion $\propto M_{\rm c}^{1/4}$, whereas our prediction contains an additional dependence on the orbit, suggesting a way to discriminate between the two scenarios. 

We show that there is a dependence of the velocity dispersion, anisotropy and rotation properties of PEs on the galactic mass profile (i.e. $\lambda$). This suggests that the PEs can be used as an independent method to determine properties of the underlying dark matter profile, which could be especially important in the core v cusp debate in dwarf galaxies (see e.g. \citealt{Walker2011}; \citealt*{Read2016}). For example, the increasing abundance of PEs with increasing $\lambda$ could lead to a higher mass-to-light ratio. Such a $\lambda$ dependent mass-to-light ratio could help explain why the metal-poor clusters Fornax 3 and Fornax 5 have an observed mass-to-light ratio higher than synthetic stellar population models (\citealt*{Larsen2012}; \citealt*{Strader2011}). 

This velocity dispersion prediction is also useful for generative models of tidal streams,  which require releasing particles from a cluster with a chosen velocity dispersion (\citealt*{Fardal2015}; \citealt*{Erkal2016}). This dispersion affects the width of the stream and therefore using the correct value is important to be able to accurately use the streams to infer galactic properties.

We compared our results to available observational data.  We used recently compiled velocity dispersion profiles from \citet{Baumgardt2017}, which contain a wide range of radial velocity and proper motion measurements from literature, and showed that most of the observed values of the velocity dispersion in the outermost bins of data lie above our prediction. There are many reasons why the observational data would increase above our prediction, including the fact that the data do not extend close enough to $r_{\rm J}$, projection effects, and that a large fraction of clusters are still under-filling their Roche volumes. Despite this, we found some clusters to be close to our prediction and not to be consistent with a prediction that would only depend on the mass of the cluster, suggesting that there is a dependence on the galactocentric distance consistent with our $\Omega^{1/3}$ dependence. With the upcoming \textit{Gaia} data it will be possible to detect stars further from the centre of globular clusters than it is currently possible. Accurately understanding the behaviour of PEs provides an independent way of inferring galactic properties and avoids the misidentification of other effects, such as the effects of a dark matter halo.

\section{Acknowledgements}
We are grateful to Justin Read, Florent Renaud, Holger Baumgardt, Douglas Heggie and Anna Lisa Varri for fruitful discussions and to the referee for useful suggestions. We are also grateful to Sverre Aarseth and Keigo Nitadori for making \texttt{\small{NBODY6}} publicly available. We also thank Mr. Dave Munro of the University of Surrey for hardware and software support.  MG acknowledges financial support from the Royal Society (University Research Fellowship), AZ acknowledges financial support from the Royal Society (Newton International Fellowship). IC, MG and AZ acknowledge support from the European Research Council (ERC-StG-335936, CLUSTERS).
\bibliographystyle{mn2e}

\label{lastpage}

\end{document}